\documentclass[journal]{vgtc}                     

\onlineid{0}

\usepackage{verbatimbox}
\usepackage{amsmath}
\usepackage{booktabs}
\usepackage{multirow}
\usepackage{graphicx}
\usepackage[utf8]{inputenc}
\usepackage[T1]{fontenc}
\vgtccategory{Research}

\vgtcpapertype{evaluation}

\title{Cybersickness, Cognition, \& Motor Skills: \\ The Effects of Music, Gender, and Gaming Experience}

\author{%
  \authororcid{Panagiotis Kourtesis}{0000-0002-2914-1064}, Rayaan Amir, Josie Linnell, \authororcid{Ferran Argelaguet}{0000-0002-6160-8015}, and \authororcid{Sarah E. MacPherson}{0000-0001-8676-6514}
}

\authorfooter{
\item
 P. Kourtesis and F. Argelaguet are with the Inria, Univ Rennes, IRISA, CNRS -- 35042 Rennes, France. E-mail: name.surname@inria.fr.
\item
R. Amir, J. Linell, and S. E. MacPherson are with the Department of Psychology, University of Edinburgh --- Edinburgh, EH8 9YL, United Kingdom. E-mail: name.surname@ed.ac.uk.
}

\shortauthortitle{Kourtesis \MakeLowercase{\textit{et al.}}: Cybersickness in VR: Cognition, Motor Skills, Music, and Gender}
\abstract{Recent research has attempted to identify methods to mitigate cybersickness and examine its aftereffects. In this direction, this paper examines the effects of cybersickness on cognitive, motor, and reading performance in VR. Also, this paper evaluates the mitigating effects of music on cybersickness, as well as the role of gender, and the computing, VR, and gaming experience of the user. This paper reports two studies. In the 1st study, 92 participants selected the music tracks considered most calming (low valence) or joyful (high valence) to be used in the 2nd study. In the 2nd study, 39 participants performed an assessment four times, once before the rides (baseline), and then once after each ride (3 rides). In each ride either Calming, or Joyful, or No Music was played. During each ride, linear and angular accelerations took place to induce cybersickness in the participants. In each assessment, while immersed in VR, the participants evaluated their cybersickness symptomatology and performed a verbal working memory task, a visuospatial working memory task, and a psychomotor task. While responding to the cybersickness questionnaire (3D UI), eye-tracking was conducted to measure reading time and pupillometry. The results showed that Joyful and Calming music substantially decreased the intensity of nausea-related symptoms. However, only Joyful music significantly decreased the overall cybersickness intensity. Importantly, cybersickness was found to decrease verbal working memory performance and pupil size. Also, it significantly decelerated psychomotor (reaction time) and reading abilities. Higher gaming experience was associated with lower cybersickness. When controlling for gaming experience, there were no significant differences between female and male participants in terms of cybersickness. The outcomes indicated the efficiency of music in mitigating cybersickness, the important role of gaming experience in cybersickness, and the significant effects of cybersickness on pupil size, cognition, psychomotor skills, and reading ability.%
}


\teaser{
  \centering
  \includegraphics[width=\linewidth]{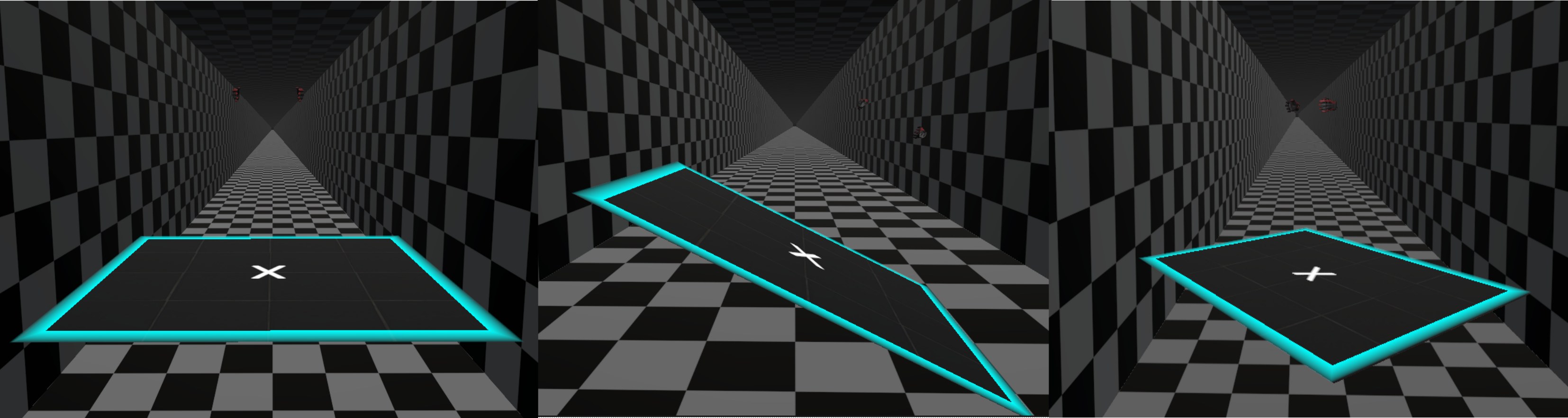}
  \caption{Examples of Accelerations in VR: Linear(Left) and Angular(Centre and Right). Note that the direction of the motion was forward.}
 \label{fig:Teaser}
 \vspace{-2mm}
}

\graphicspath{{figs/}{figures/}{pictures/}{images/}{./}} 

\usepackage{tabu}                      
\usepackage{booktabs}                  
\usepackage{lipsum}                    
\usepackage{mwe}                       

\usepackage{mathptmx}                  

\begin{document}


\maketitle
\section{Introduction}
Virtual Reality (VR) has been implemented in many areas such as education \cite{Radianti2020}, professional training \cite{Xie2021}, cognitive assessment \cite{Kourtesis2021}, mental health therapy \cite{Emmelkamp2021}, and entertainment \cite{Cruz-Neira2018}. However, a principal drawback of VR is the presence of cybersickness that affects a percentage of the users \cite{Rebenitsch2021}. Cybersickness pertains to nausea, disorientation, and oculomotor symptoms. While cybersickness has similarities with simulator sickness, it differs to simulator sickness in terms of severity, with accounts reporting increased general discomfort due to nausea- and disorientation-related symptoms \cite{Stanney1997}. Furthermore, cybersickness differs from motion sickness as cybersickness is triggered by visual stimulation and not by actual movement \cite{Davis2014}.%

While there is not a comprehensive theory of cybersickness, the predominant one is the sensory conflict theory, which postulates that cybersickness symptomatology is induced by a conflict between our vestibular (inner ear) and visual system \cite{Rebenitsch2021,LaViola2000}. In these terms, while motion is perceived visually, the inner ear (body) does not perceive motion, which results in a conflict between proprioception and vision. As such, the vection, an illusory sense of motion, that occurs in VR is the predominant reason for experiencing cybersickness \cite{Nesbitt2017,Kim2021}. In VR, linear and angular accelerations may occur (see \autoref{fig:Teaser}) that induce cybersickness in the user.%

Cybersickness may affect the cognitive and/or the motor performance of the user. However, there are discrepant results in the relevant literature. For example, Dahlman et al. \cite{Dahlman2009} found that cybersickness  significantly decreases verbal working memory ability. In contrast, Mittelstaedt et al. \cite{Mittelstaedt2019} found no significant decrease in visual working memory. Nonetheless, several studies have found a substantial decrease in reaction times \cite{Mittelstaedt2019,Nesbitt2017,Nalivaiko2015}. Thus, while there are inconclusive results regarding cybersickness's effects on cognition, there is agreement regarding its effect on psychomotor skills.

Furthermore, individual differences may play a role in how cybersickness is experienced. For example, the gender of the user has been suggested as a factor that could influence the intensity of cybersickness \cite{Stanney2020,Stanney2003}. The reasons were attributed to differences in hormones and interpupillary distance (IPD) between females and males \cite{Stanney2020,Stanney2003}. However, there are mixed reports in the literature, where some studies report higher cybersickness in female participants, and others an absence of significant differences between males and females \cite{Saredakis2020}. Also, other factors such as VR, computing, and/or gaming experience may modulate the user's resilience or susceptibility to cybersickness \cite{Saredakis2020}. Yet, the relationship between these aspects and cybersickness have not been adequately examined \cite{Saredakis2020}.%

Towards the mitigation of cybersickness symptoms, several approaches have been used, albeit that each has their limitations. For instance, “adaptation” (i.e., exposing an individual repeatedly to the cybersickness stimuli assists with building a tolerance to it) may be an efficient way to minimize cybersickness\cite{Cheung2005}. However, this process is time-consuming, costly, and requires considerable commitment. Moreover, anti-sickness drugs and supplements (e.g., ginger) have been suggested \cite{Wood1988,Golding2015}. Yet, these are invasive approaches, and they may also cause other side effects (e.g., drowsiness) or allergic reactions \cite{Golding2015}. In HCI, several techniques have been proposed such as the viewpoint control \cite{Farmani2020}, dynamic blurring \cite{Nie2020}, postural stability \cite{Risi2019}, and taking short breaks \cite{Szpak2019}. Nonetheless, these approaches prevent certain interactions with the virtual environment (e.g., viewpoint control and postural stability) and/or may negatively affect the user experience (e.g., blurring and short breaks).%

Alternatively, in medical settings, calming music has already shown promising results in mitigating nausea-related symptomatology \cite{Karagozoglu2013}. Similar results have been found in mitigating motion sickness, where calming music was found to substantially decrease the intensity of motion sickness  \cite{Sang2006}. Furthermore, pleasant music was able to alleviate motion sickness during the simulation of a bike ride \cite{Keshavarz2014}. In a similar study, the liked music was also found to alleviate motion sickness during the simulation of a bike ride \cite{Peck2020}. These promising results indicate that music may offer relaxation (e.g., an effect against stress) or a distraction (e.g., favourite or pleasant music prevents the fixation on stimuli that induce or worsen motion sickness), which mitigate the symptomatology of motion sickness. However, as far as we are aware, no study has yet examined the effects of music on cybersickness, especially in immersive VR.%

This paper reports two studies, where the 1$st$ study (N = 92) selects music tracks that are universally rated as calming or joyful, and the 2$nd$ study (N = 39) offers an extensive examination of cybersickness in VR. Specifically, the mitigating effects of calming and pleasant music on cybersickness are assessed. In addition, the effects of cybersickness on cognition and motor skills, as well as reading ability and pupil size, are evaluated. Finally, the role of demographics such as gender and VR, computing, and gaming experience are examined. The contributions of this paper can hence be summarized as follows:%
\begin{itemize}%
    \item
    Providing evidence that pleasant and calming music mitigate nausea-related symptoms in VR, and pleasant music substantially alleviates overall cybersickness symptomatology.%
    \item
    Highlighting the effects of cybersickness effects on cognition, motor skills, reading ability, and pupil size.%
    \item
     Providing evidence that users with higher gaming experience are more resilient to cybersickness, while users with lower gaming experience are more susceptible to it.%
    \item
    Examining gender differences in terms of cybersickness by evaluating and considering the important role of gaming experience on cybersickness.%
\end{itemize}%
\section{Relevant Work}
\subsection{Effect of Music on Motion Sickness}
Music has been previously been implemented to alleviate motion sickness. In medical applications, there is strong evidence that relaxing music decreases nausea-related symptomatology (e.g., due to chemotherapy) \cite{Karagozoglu2013}. In a non-medical setting, Sang et al. \cite{Sang2006} were among the first to demonstrate that music may mitigate motion sickness. Participants (N = 24) were placed in a simulator designed to trigger motion sickness, through body rotations and head movements, for 30 minutes. The Simulator Sickness Questionnaire (SSQ) was used to evaluate sickness levels before and after the simulation. While this study provides insight into the benefits of music on motion sickness, many questions remain unanswered. Firstly, the music was played after motion sickness symptoms appeared rather than during the simulation, possibly altering the results. Also, a single music track was used, without providing a clear rationale for this selection, and they did not consider different types of music and their possible influence on motion sickness. Finally, the study was on motion sickness (i.e., body movements), but cybersickness was not studied (i.e., visually induced).%

Similarly, Keshavarz and Hecht \cite{Keshavarz2014} investigated the role of pleasant music on motion sickness. Participants (N = 93) were shown a first-person video (14 mins) of a bicycle ride with shaky movements on a large screen. Participants were randomly assigned to four groups: 1) relaxing; 2) neutral music; 3) stressful music; or 4) no music. During the video, the Fast Motion Sickness Scale (FMS) was used to evaluate motion sickness. The SSQ was also administered before and after the video. Participants also rated the pleasantness of the music. Regardless of the condition, music rated as pleasant significantly lowered motion sickness levels, whereas music rated as unpleasant did not. While these are promising results, the study had a between subjects design. Given that motion sickness varies significantly across individuals \cite{Golding2015,Risi2019}, a within design would have been more appropriate. Notably, the study was not conducted in immersive VR, where vection is substantially stronger, and consequently so are the visually induced symptoms \cite{Dennison2016}.%

In a follow-up study, Peck et al. \cite{Peck2020} examined how valence, arousal, and the liking of music affect motion sickness. Participants (N = 80) were allocated to one of four groups, where music of different valence and arousal was presented (i.e., happy, peaceful, agitated, and sad). Participants evaluated valence and arousal, and opted for their favourite music.  Using the same video as above (see \cite{Keshavarz2014}), four groups (N = 20 for each) were formed, where the first was presented with their favourite music, the second had high valence music, the third had high arousal music, and the control group had no music. The results showed an effect only for the highly liked music. However, this study has the same limitations as Keshavarz and Hecht, where a within design would be more efficient for controlling interindividuality in motion sickness, and immersive VR would offer more intense vection and symptomatology.%

\subsection{Effects of Cybersickness on Cognitive and Motor Skills}
Cybersickness has been found to substantially decrease the cognitive and motor performance of the user in immersive VR \cite{Saredakis2020}. Nevertheless, the reports are not consistent. In favour of temporary cognitive decline, the study of Dahlman et al. \cite{Dahlman2009} postulated that motion sickness significantly decreases the verbal working memory of users. In this study (N = 38), participants were seated on a rotating optokinetic drum. Motion sickness and verbal working memory were examined before and after the session. The participants who experienced motion sickness had severely impaired verbal working memory in the post-session assessment. In contrast, the participants with low motion sickness did not show any decrease in verbal working memory. While the study supports cognitive decline due to motion sickness, the experiment is strongly related to simulator studies (i.e., rotating drum), rather than the visually induced symptomatology observed in immersive VR.%

In immersive VR, Varmaghani et al. \cite{Varmaghani} conducted a study (N = 47) where the participants were allocated to two groups, a VR group (N = 25) or a control group (N = 22; playing a board game). Cybersickness and cognition were assessed before and after the respective session. The Corsi blocks task (visuospatial memory), Manikin spatial task (visuospatial processing), and Color Trails test (attentional processing and shifting) were administered. A priming effect was observed in the Manikin and Color Trails test. The outcomes showed that the VR group did not increase its visuospatial processing ability, while the control group did. This result indicated an effect of cybersickness on visuospatial processing or learning ability. However, in the remaining measures, no differences were observed. Also, the cybersickness scores did not correlate cognitive performance. Therefore, this study provided mixed results. However, they did not consider verbal memory as in \cite{Dahlman2009}. In addition, while cybersickness varies across individuals, this study had a between subjects design which does not permit consideration of random effects due to individual differences. Also, while an absence of correlations was reported, correlations do not provide information regarding the causal or predictive ability of a variable (e.g., cybersickness) towards another variable (e.g., cognition) \cite{Freedman}. To examine predictions and/or causation, a regression analysis (models) would have been required\cite{Freedman}.%

In another study, Mittelstaedt et al. \cite{Mittelstaedt2019} examined cybersickness and cognition in pre- and post-sessions in VR. The participants (N = 80) were allocated into one of four groups (N = 20 for each), where different input devices or displays were used in the three groups (i.e., Gamepad-HMD, Bike-HMD, Bike-Screen), and a control group (i.e., resting in between). The SSQ (cybersicknes), the Deary–Liewald Reaction Time task (reaction time), the Mental Rotation task (spatial processing), the Corsi Block task (visuospatial working memory), the Arrow task (reaction time), and the Visual Search task (visual attention processing) were used. The reaction times of the VR groups were poorer on both corresponding tasks. Also, the VR groups did not show the improvement of the control group on the attentional processing task. On spatial processing and visuospatial working memory, there were not significant differences among groups. Overall, these results suggest a negative effect on attentional processing and reaction times, while spatial abilities and visuospatial memory remain intact. While the results support an effect of cybersickness on cognition, the outcomes were not robust. Similar to the aforementioned studies, this study had a between subjects design that does not allow for consideration of individual differences.%

Furthermore, the studies of Nalivaiko et al. \cite{Nalivaiko2015} (N = 26) and Nesbitt et al. \cite{Nesbitt2017} (N = 24) examined the effects of cybersickness on reaction times. The cybersickness and reaction times were evaluated before and after exposure. The SSQ (cybersickness) and the Deary–Liewald Reaction Time task were used. The VR session consisted of a roller coaster ride, where substantial linear and angular accelerations took place to induce cybersickness in participants. In both studies, the reaction times were significantly decelerated in the post-exposure assessments. Furthermore, both studies reported significant correlations between the slowing of reaction times and the increased intensity of nausea-related symptoms. Taking together the results of \cite{Nalivaiko2015,Nesbitt2017,Mittelstaedt2019}, there is substantial evidence that reaction times are negatively affected by cybersickness.%

However, none of these studies measured cybersickness during VR exposure. Furthermore, none of the studies evaluated the reaction times and cognition during the VR session (i.e., during immersion) which raises questions such as whether there is an effect only after exposure or also during exposure. Finally, while the study of Mittelstaedt et al. \cite{Mittelstaedt2019} attempted mixed regression models to examine the effects of cybersickness on cognition and reaction times, the studies of Nalivaiko et al. \cite{Nalivaiko2015} (N = 26) and Nesbitt et al. \cite{Nesbitt2017} only explored correlations between cybersickness and reaction times. Thus, further investigation is required to appraise the effects of cybersickness on reaction times and cognition.%

\subsection{Gender and VR/Computing/Gaming Experience}
Individual differences appear to play a significant role on how and when cybersickness is experienced \cite{Saredakis2020}. The most common aspect that has been examined is the gender of the user, where female users experience substantially more severe cybersickness than male users. Yet, the results are inconsistent across studies. For instance, Petri et al. \cite{Petri2020} conducted a study (N = 30; 15 females and 15 males) to explore gender differences. In a seated position, the participants had to observe a karate demonstration while immersed in VR. While no significant differences in objective metrics (i.e., heart rates and skin conductance) was observed between female and male participants, the subjective metrics (SSQ) revealed a significant difference between female and male participants. In contrast, the study of Melo et al. \cite{Melo2018} showed no significant difference between female and male participants. However, in this study, cybersickness intensity was not high, since there were no accelerations in the virtual environment.%

In contrast, in the study of Stanney et al. \cite{Stanney2020}, participants (N = 46) were allocated into two groups: a VR environment (N = 30; 15 male/15 female) or a flat computer screen (N =. 16; 8 male/8 female). Cybersickness was measured before and after exposure. Gender differences were detected only in the flat screen condition. In the 2$nd$ experiment (N = 120), the participants were allocated into eight groups (N = 15 per group) defined by gender (male vs. female), IPD (fit vs. non-fit), and motion sickness history (low vs. high). In this experiment, gender was not a significant predictor of  cybersickness. Instead, only the IPD fit, motion sickness history, exposure duration, and EGG bradygastria were significant predictors. However, there were a great number of variables in the model (12 variables), which may have masked the effects of other variables on cybersickness, such as gender, gaming experience, VR exposure, hormonological cycle, and anxiety.%

Furthermore, a meta-analysis of studies on cybersickness revealed no significant differences between male and female users \cite{Saredakis2020}. The meta-analysis considered only studies that were conducted in VR, and where the SSQ was administered for evaluating cybersickness. While this outcome indicates an absence of differences, the included studies were specific to cybersickness, hence, the cybersickness levels were adequately varied between them. Nevertheless, the authors also speculated that other factors should be considered such as VR and gaming experience. However, the past VR studies on cybersickness did not consider these aspects \cite{Saredakis2020}.%

To our knowledge, only three cybersickness  studies exist that examine computing, VR, or gaming experience. The Stanney et al. \cite{Stanney2020} study considered gaming experience but did not find an effect of gaming experience on cybersickness. However, the measurement of gaming experience in this study is unclear. Traditionally, gaming (or any IT) experience is calculated by considering both the ability (i.e., how well the user operates something) and the frequency (i.e., how often the user operates something) \cite{Kourtesis2019, Smith}. In Stanney et al. \cite{Stanney2020}, a description evaluating gaming experience is not explicitly provided, indicating that the authors did not focus on this variable.%

Moreover, the study of Kourtesis et al. \cite{Kourtesis2019} reported no effect of gaming or VR experience on cybersickness. However, this study only reported very low intensities of cybersickness since commercial VR games were used. On the other hand, the study of Weech et al. \cite{Weech2020} explored the effect of gaming experience on cybersickness. In this large study (N = 153), participants were allocated into two groups: 1) an exploration with a narrative, and 2) an exploration without a narrative. The results indicated that the narrative was effective in mitigating cybersickness, however, this was effective only when the users had higher gaming experience. Overall, this study suggested that gaming experience has an (direct or indirect) effect on cybersickness.%

\subsection{Summary of Literature Review}
\label{ssec:Review}
\paragraph{Music Effect on Cybersickness:}Based on the literature review discussed above, only the effects of music on motion sickness have been examined. There is hence a gap in the literature regarding the effects of music on cybersickness in immersive VR, where vection and symptoms are stronger. Also, the studies on motion sickness and music did not have a within subjects design, which prevented them from considering variability (random effects) across individuals. Finally, only pleasant (high valence) music has been used in previous studies, while calming music has shown promising results in clinical settings.%
\vspace{-0.5mm}
\paragraph{Cybersickness's Effects on Cognitive and Motor Skills:} The literature review showed an agreement regarding the effects of cybersickness on reaction times (i.e., psychomotor skills). Visuospatial ability and visuospatial working memory appear not to be affected by cybersickness. However, while verbal working was found to be affected by motion sickness, there is not a VR study investigating cybersickness's effect on it. Verbal working memory is crucial for encoding new memories/information and facilitating the fine functioning of high order cognitive abilities like episodic memory, language, and learning \cite{Cowan}. Given that there are VR applications in professional training, education, and therapies, the examination of cybersickness's effects on verbal working memory is essential. Finally, studies have methodological shortcomings. They did not assess cybersickness and cognition during VR exposure/immersion, and the effects were mainly examined via correlations instead of mixed regression models.%
\vspace{-0.5mm}
\paragraph{Gender and VR/Computing/Gaming Experience:} The literature review revealed contradictory outcomes regarding gender differences in terms of cybersickness. While a meticulous study indicated a gender effect on cybersickness (mainly due to the IPD), the meta-analysis showed no significant differences. The role of gender has to be further examined by considering other factors that may modulate cybersickness. Finally, the role of gaming, VR, and/or computing experience is under-investigated, while the there is a suggestion that gaming may have an effect on cybersickness.%
\vspace{-0.5mm}
\paragraph{Conclusions:}In conclusion, the current study will offer the first examination of music as a mitigating technique for cybersickness in VR. In addition, this study will be the first to study the effect of cybersickness on verbal working memory. Similarly, this study will also be the first to assess gender differences in terms of cybersickness, while it also considers the effects of VR, computing, and gaming experience on cybersickness.  Finally, the current study will be the first to examine cognition and cybersickness in during VR exposure/immersion, while considering for variability across individuals, by developing and implementing a 3D-VR version of a cybersickness questionnaire.%
\vspace{-0.5mm}
\section{Online Study: Evaluation of Calming and Joyful Music Tracks}
An online questionnaire study was conducted to obtain ratings and rankings of joyful and calming music tracks. The rating and rankings assisted with selecting two music tracks most perceived as joyful or calming. The identified best tracks were used in our main VR study.%

A total of 92 participants (51 males, 39 females, 2 non-binary) took part in the online questionnaire study. Participants were approached via social media. Approval for this online questionnaire study was granted by the School of Philosophy, Psychology, and Language Sciences (PPLS) Research Ethics Committee of the University of Edinburgh. All participants provided their online consent before beginning the survey. A total of twelve pre-selected instrumental tracks (each lasting approximately 30 seconds) were used. Music tracks were selected based on previous literature or were similar to those used in prior studies (e.g., \cite{Melo2018,STANGLMEIER,Tso,Graff796}).%

Firstly, participants rated the first category of tracks (i.e., calming), which consisted of six calming tracks. A 7-point Likert scale, where 1 indicated not at all calming and 7 indicated extremely calming, was presented to participants, asking them to rate how calming the track was. Participants were instructed to listen to the whole 30 second clip before submitting a response, and were unable to pause the clip or proceed onto the next track prior to the clip's completion. Following the individual ratings, a list of each of the tracks (30 seconds each) was presented to each participant. The participants were instructed to listen to each clip once more and rate these in comparison to one another. Participants were asked to drag and drop (rank) the same six tracks in order of preference.  Following this, the same procedure was repeated for the six joyful tracks, with the rating of individual tracks followed by an overall comparative ranking.%

The joyful track included the the Sock Shopping by Ray Amir (\emph{Joyful A}), the Brandenburg Concerto No.4 in G Major, Allegro by Bach (\emph{Joyful B}), the Good Times (instrumental version) by Chic (\emph{Joyful C}),  the Thunder and Lightning polka by Johann Strauss (\emph{Joyful D}), the Spring by Vivaldi (\emph{Joyful E}), and the Summertime by Morning Light Music (\emph{Joyful F}). The calming tracks included the Watermark by Enya (\emph{Calming A}), the Piano Bar Jazz by Pinegroove Music (\emph{Calming B}), the Mellow Sky by Ray Amir (\emph{Calming C}), the Maharishi Gandharva Veda by Amar Nath (Calming D), the Electra by Airstream (\emph{Calming E}), and the  Weightless by Marconi Union (\emph{Calming F}). The 30 seconds excerpts of the music tracks can be downloaded from the Open Science Framework (OSF) repository through this \href{https://osf.io/qh6wd}{LINK}%
\subsection{Results}
Analyses were conducted in R \cite{R}. The final track selection was determined by considering both rating and ranking. Wilcoxon signed-rank tests were performed to compare tracks within categories, and the p-values were corrected for multiple comparisons.%
\paragraph{Joyful Tracks:}The Wilcoxon signed-rank test showed tracks B, C, and E to have significantly higher ratings than the other tracks ($p < .05$). However, there was no significant difference among these tracks. The comparisons of rankings of these tracks showed that track C had a significantly better ranking than B ($p < .05$), while E did not differ substantially. According to the frequencies of the tracks' ratings, 82.6\% of the respondents rated Joyful C with a 4 (i.e., joyful) or higher. Notably, 59.8\% of respondents gave track C a rating of 5 (very joyful) or greater (i.e., very much joyful or extremely joyful), the highest of all tracks. Therefore, Joyful track C (i.e., the instrumental version of Good Times  by Chic) was selected for the VR experiment. The two best joyful tracks can be downloaded from the OSF repository through this \href{https://osf.io/vf8sw}{LINK}
\vspace{-1.5mm}
\paragraph{Calming Tracks:}The Wilcoxon signed-rank test showed tracks C and F to have significantly higher ratings than the other tracks ($p < .05$). However, the comparison of rankings indicated that track C has a significantly better ranking than the others ($p < .05$). Additionally, according to the frequencies of the tracks' ratings, 87.7\% of the respondents rated Calming C with a 4 (i.e., calming) or higher. Importantly, 64.1\% of the respondents gave track C a rating of 5 (i.e., very calming) or greater (i.e., very much calming or extremely calming), the highest of all tracks. Thus, Calming track C (i.e., the Mellow Sky by Ray Amir) was chosen for the VR experiment. The two best calming tracks can be downloaded from the OSF repository through this \href{https://osf.io/s6yqc}{LINK}
\vspace{-1.5mm}

\section{Immersive VR Study}
The main experiment was conducted to examine the effects of cybersickness on cognition (i.e., verbal working memory and visuospatial working memory) and motor skills (i.e., reaction  time). Equally, this experiment strove to examine the effects of music, gender, and IT experience (VR, gaming, and computing experience) on cybersickness.%
\subsection{Virtual Environment Design and Interactions}
The Unity3D game engine was used for the development of the virtual environment. The SteamVR SDK was used for designing the interactions with it. Since gaming experience may modulate task performance  \cite{Kourtesis2021}, the virtual hands/gloves of SteamVR SDK were used for ensuring an ergonomic and effortless interaction. Importantly, the interactions did not require any button to be pressed, everything was facilitated by touching the object (initial selection) and a continuous touching of the object (a confirmation of the selection). Also, the SteamVR virtual gloves did not connote any gender or race, which assisted with preventing confounding effects from these variables \cite{Schwind2017}.%

Furthermore, Amazon Polly was used to produce audio clips with neutral naturalistic voices, to provide the instructions to the users. Note that users received instructions in video, audio, and written form, to ensure their understanding and seamless completion of the tasks. The audio (especially for feedback sounds) was spatialized using the SteamAudio plugin. The SRapinal SDK was used for eye-tracking and facilitating the pupillometry, as well as the fixation duration over the text of the cybersickness questionnaire (i.e., reading ability). Finally, the randomization of the experimental blocks within and between participants, extraction of the data into a CSV file, and the experimental design and control were attained using bmlTUX SDK \cite{bmlTUX}.%

\subsubsection{Rides with Linear and Angular Accelerations/}
\label{ssec:Rides}
As the literature suggests, linear and angular accelerations are able to induce significant cybersickness symptoms to the users in a relatively short time (e.g., $\sim$ 5 - 10 mins) \cite{Risi2019,Nalivaiko2015,Nesbitt2017,Saredakis2020,Kourtesis2019}. In accordance, we designed the ride to last 5 mins, since it had to be repeated three times (i.e., total 15 min ride) for each participant, and it had to last comparably with the duration of the music tracks ($\sim$  5 mins per track). The ride was designed as an animation of the platform that the user was standing on (see \autoref{fig:Teaser}). The overall direction of the motion was always forward (except the last stage; see reversed z axis). The movements were similar to those of a roller coaster. The ride consisted of the following accelerations (in this specific order): a) linear (z axis), angular (z and y axes), angular (z, x, and y axes), angular (roll axis), extreme linear (z axis), angular (yaw axis), and extreme linear (y axis followed by reversed z axis). The environment was a simple black and white surrounding (see \autoref{fig:Teaser}) to ensure that the symptoms were only induced by vection, and not due to other reasons (e.g., intense colours). The squared/tiled design assisted with providing cues for eliciting vection and perceiving the altitude changes.%

\subsection{Cognitive and Psychomotor Skills' Assessment}
\label{ssec:Tests}
Since the examination of cybersickness, cognition, and motor skills had to be repetitive, while the user is immersed in VR, immersive VR versions of well-established tasks/tests were developed. For the development of these VR cognitive and psychomotor tasks, the specific design and development guidelines and recommendations for cognitive assessments in immersive VR were followed \cite{Kourtesis2019b}.%

\subsubsection{Verbal Working Memory}
A VR version of the Backward Digit Span Task (BDST; \cite{wechsler}) was developed and used. In the VR BDST, the participants listened to a series of digits, which they needed to remember and recall in the reverse order of their presentation. For example, if participants heard 2, 4, and 3, then they should respond by indicating the reverse order (i.e., 3, 4, 2). After listening to the digits, a keypad appeared in front of participants, which they had to use for providing the digits in the reverse order. To indicate a number, participants touched the white box button that displays the equivalent number (see \autoref{fig:Tests}). While touching the button, it turned blue. To confirm their response, participants needed to keep touching the button for a second. Then, if the response was correct, the button turned green, and made a positive sound. If the response was incorrect, then the button turned orange, and made a negative sound. Each trial ended when participants made a mistake or when they provided the digits correctly in their reverse order. Every second successful trial, the length of the digit sequence was increased. The task was ended when they made two subsequent mistakes within the same digit sequence length (e.g., 3 digits), or when they finished the last trial (i.e., second trial with a sequence length of 7 digits). The \emph{Total Score} was the addition of the number of successful trials and the highest digit sequence length that at least one trial was successfully performed.%

\subsubsection{Visuospatial Working Memory}
\begin{figure}[!h]
 \centering 
 \includegraphics[width=\columnwidth, height = 180pt]{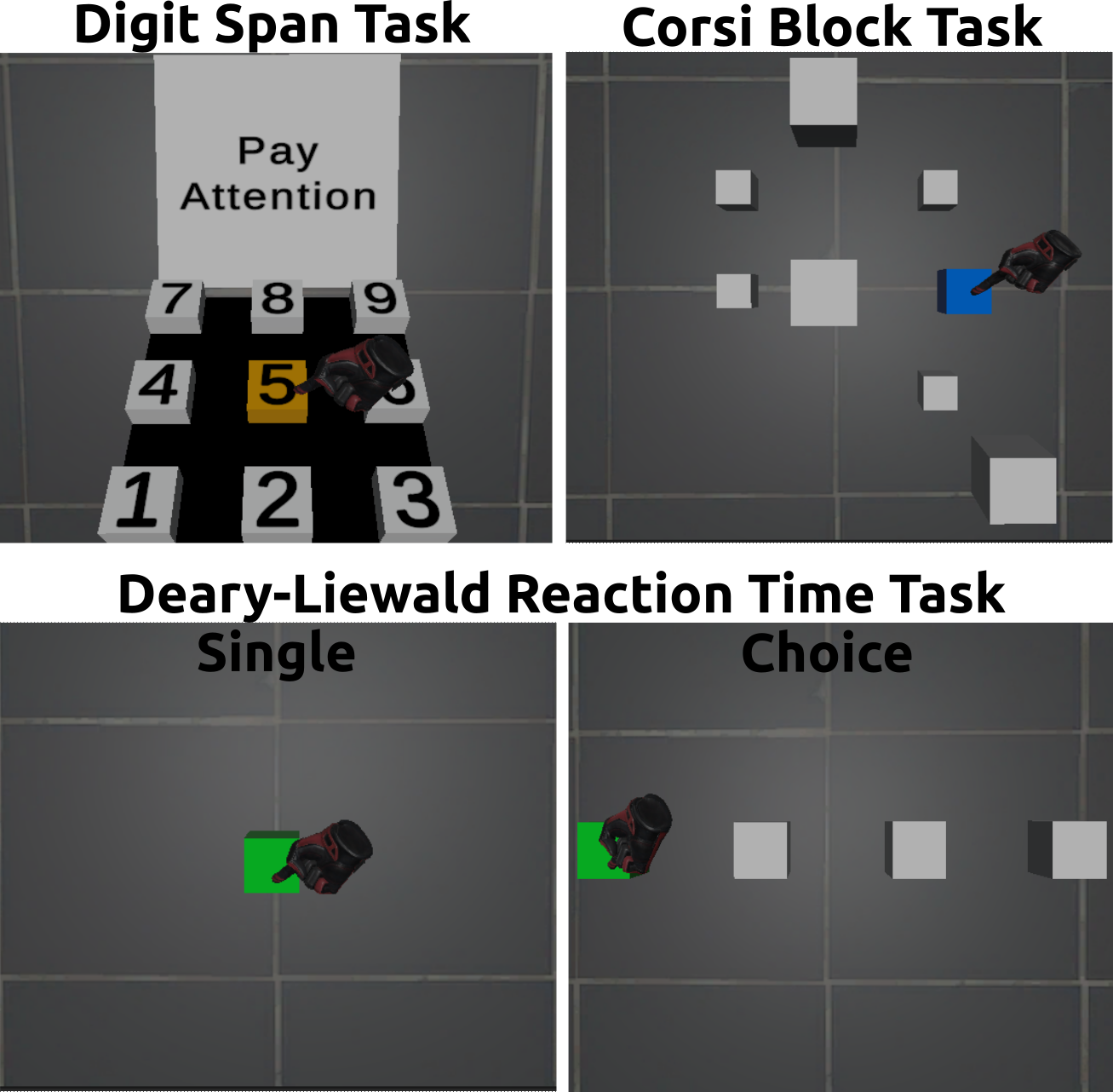}
 \caption{Digit Span Task (Upper Left), Corsi Block Task (Upper Right), and Deary Liewald Reaction Time Tasks (Bottom).}
 \label{fig:Tests}
 \vspace{-5mm}
\end{figure}

Visuospatial working memory was assessed using the VR version of the Backward Corsi Block Test (BCBT)  \cite{Corsi} that we developed. In this task, there are 27 white boxes, each one is placed in a different position on the x, y, and z axes. However, out of the 27 possible boxes, participants were randomly shown only nine boxes in each trial (see \autoref{fig:Tests}). Each trial began with the presentation of 9 boxes. A number of these boxes (depending on the current sequence length) were randomly presented (turning blue and making a bell sound) in  sequential order. Each box from the sequence was presented for one second. At the end of the sequence's presentation, participants had to select the boxes in the reverse order. To select a cube, participants touched it and it turned blue. After touching the cube for one second, the target was selected. Once selected, it either turned green and made a positive sound (i.e., correct response), or it turned orange and made a negative sound (i.e., an error). The trial was ended when participants either made a mistake, or they correctly selected all the targets in their reverse order. The sequence lengths were initially two boxes, with two attempts for each length. The number of boxes in the sequence was increased by one box when at least one of the two trials of the same length/span was correct. When the participant incorrectly recalled two sequences of the same length, the task was ended. Likewise, when the second trial of the last length/span (i.e., 7 cubes) was performed, then the task was ended. Sequence lengths increased up to seven which indicated a perfect score if no mistakes were made. The \emph{Total Score} was calculated by summing the span (correct sequence with the most cubes) and the total number of correct sequences.%

\subsubsection{Psychomotor Skills}
A VR version of the Deary–Liewald (DL) Reaction Time task \cite{Deary} was developed and used to assess reaction times. The DL Reaction Time Task (also referred to as the DL task) contained two tasks for assessing simple reaction time (SRT) and choice reaction time (CRT). During the SRT task, participants were asked to observe a white box and to touch it as soon as the box changed into blue (see \autoref{fig:Tests}). The SRT task consisted of 20 trials/repetitions. In the CRT task, one of the four boxes (aligned horizontally) randomly changed blue. Participants had to touch the box, once the box turned blue (see \autoref{fig:Tests}). This task consisted of 40 trials/repetitions. In both SRT and CRT, the participants were instructed to touch the boxes as fast as possible using either hand. Both the SRT and CRT had a practice session at the start to ensure that the instructions were understood by the participants.%

The SRT was scored by averaging the total reaction time across the 20 trials. Likewise, the recorded times for the CRT were averaged to produce a score. However, the CRT had three scores. The eye-tracking enabled us to measure the time required to attend to the target, since the time it was presented (Attentional Time). Also, this allowed us to calculate the time required to touch the target since the time that was attended (Motor Time). Finally, the overall time between the target's presentation until its selection (Reaction Time) was also measured. Thus, three scores were produced: 1) the \emph{Reaction Time (RT)} to indicate the overall psychomotor speed; 2) the \emph{Attentional Time (AT)} to indicate the attentional processing speed; 3) the \emph{Motor Time (MT)} to indicate the movement speed.%

\subsection{Cybersickness in VR Questionnaire}
\begin{figure}[!h]
 \centering 
 \includegraphics[width=\columnwidth,height = 150pt]{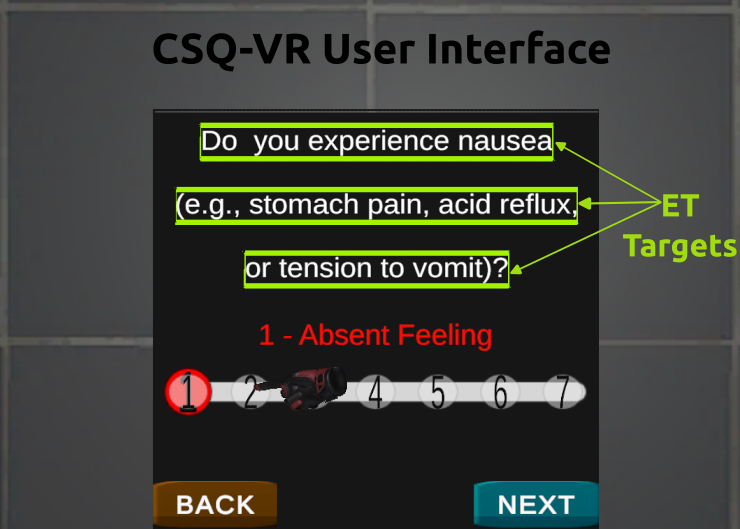}
 \caption{CSQ-VR User Interface and Eye-Tracking (ET) Targets.\\ Note: Eye-tracking targets were not visible to the user.}
 \label{fig:CSQ-VR}
 \vspace{-6mm}
\end{figure}

The Cybersickness in VR Questionnaire (CSQ-VR) was used to assess cybersickness symptomatology and intensity. The CSQ-VR is derived from the VR Induced Symptoms and Effects of the VR Neuroscience Questionnaire, which has been found to have very good structural and construct validity \cite{Kourtesis2019}. The CSQ-VR has been validated against the SSQ and the VRSQ, and showed superior reliability to them \cite{KourtesisCSQ}. The advantages of CSQ-VR pertain to its short administration procedure (only 6 items/questions) and the production of easily comprehensible outcomes \cite{KourtesisCSQ,Somrak}. Moreover, the CSQ-VR assesses all the sub-types of cybersickness symptoms, such as nausea, disorientation, and oculomotor. For each sub-type, there are 2 questions. Each question is presented on a 7-item Likert Scale, ranging from \emph{"1 - absent feeling"} to \emph{"7 - extreme feeling"}, where each response is a combined text and number (see \autoref{fig:CSQ-VR}). The CSQ-VR produces a \emph{Total Score} and three subscores, one for each type of symptom (i.e., a \emph{Nausea Score}, a \emph{Disorientation Score}, and a \emph{Oculomotor Score}). Each subscore is calculated by adding the two corresponding responses, and the total score by adding the three subscores. The CSQ-VR can be accessed via this  \href{https://www.researchgate.net/publication/366400274_CyberSickness_in_Virtual_Reality_Questionnaire_CSQ-VR_A_brief_tool_for_evaluating_the_Virtual_Reality_Induced_Symptoms_and_Effects_VRISE}{LINK}.%

Since the aims of this study required the repeated assessment of cybersickness while the user is immersed in VR, the use a short, yet inclusive and valid, questionnaire like CSQ-VR was preferred. An User Interface (UI) was designed and developed where the question is displayed in the upper region, and the response (in red) in the middle region. Using a slider, the users could change their response, by touching the corresponding number, or sliding along the slider (see \autoref{fig:CSQ-VR}). In addition, based on the established link between verbal working memory and reading speed \cite{Dixon}, as well as between pupil size and affective/emotional state \cite{Partala}, we integrated eye-tracking metrics to explore these aspects. Invisible eye-tracking targets were placed in front of the text, while their height and width was always matched to the displayed text per line (see \autoref{fig:CSQ-VR}). The use of these targets allowed us to measure the \emph{Fixation Duration} over the text as a metric of reading speed. Also, the measurement of pupil size was continuous while the user  responded to the CSQ-VR questions. This enabled us to measure the average \emph{Pupil Size} (right and left) during this process, as a physiological metric of negative emotion.%

\subsection{Participants and Procedures}
An a priori power analysis indicated a sample size of 36 participants or greater is required for the statistical analyses. A total of 39 participants (22 females, 17 males) aged between 22 and 36 years [$M(SD) = 25.28(3.25)$] were recruited for the current study. Education (in years) ranged from 13 to 20 years [$M(SD) = 17.23(1.60)$]. For the comparisons between male and female participants, for being matched in terms of age and education, a sample of 32 (16F and 16M)  was considered. Recruitment was achieved via opportunity sampling, using the internal mailing lists of the University of Edinburgh, alongside advertisements on social media. The study was approved by the School of PPLS Ethics Committee of the University of Edinburgh. Written consent was obtained from all participants prior to their participation. Participants were compensated with £20 each for their time and effort.%

The Motion Sickness Susceptibility Questionnaire (MSSQ;\cite{GOLDING}) was completed prior to enrolment, to reduce the likelihood of severe symptoms following VR exposure. In line with the MSSQ author suggestions \cite{GOLDING}, the 75th percentile was used as a parsimonious cut-off score for inclusion in the study. This allowed us to exclude the individuals who are prone to experience strong symptoms (i.e., the upper 25th percentile of the population). The included participants were then invited to attend the experiment. Upon arrival, participants were informed of the study's aims and procedures, and the adverse effects that they may experience. The participants then provided their informed consent.%

Every participant went through an induction on how to wear the headset and use and hold the controllers. An HTC Vive Pro Eye was used, which embeds an eye-tracker with a binocular gaze data output frequency of 120Hz (i.e., refresh rate), a tracking accuracy of 0.5°-1.1°, a 5-point calibration, and a 110° trackable field of view. The participants then provided their demographic data: age, sex, gender, education, dominant eye, VR experience, computing experience, and gaming experience, by responding to a questionnaire.VR/computing/gaming experience were calculated by adding scores from two questions (6-item Likert Scale) for each one. The first question was regarding the participants' ability (e.g., 5 - highly skilled) to operate a VR/computer/game, and the second one was regarding the frequency of operating them (e.g., 4 - once a week).%

After the demographics, participants were immersed in VR. Note that during immersion, the participants were always (i.e., assessments and rides) in a standing position in the middle of the VR area (see X mark in \autoref{fig:Teaser}. They started with the baseline assessment, during which a video tutorial was offered, alongside verbal and written instructions, prior to each questionnaire or task. Also, since previous research indicated that the subjective perception of the music track may affect the efficiency of the music in mitigating motion sickness \cite{Peck2020}, we included a questionnaire (only in the baseline condition) to obtain participants' ratings. Using an UI comparable to CSQ-VR (see \autoref{fig:CSQ-VR}), participants listened to an excerpt (30 secs) of each track (Calming and Joyful), and then rated them based on a 7- point Likert scale (1 = joyful/calming; 7 = extremely joyful/calming).%

The rest of the tasks were performed again after each ride: the CSQ-VR questions, the verbal working memory task (BDST), the visuospatial working memory task (BCBT), and the reaction time task (DL; see \autoref{ssec:Tests}). After the baseline assessment, the first ride started. In each ride (see \autoref{ssec:Rides}), a different type of music was offered (i.e., Joyful, Calming, or No Music), where each track lasted approximately 5 minutes (i.e., the full track), throughout the ride. The participants completed four assessments (Baseline + one assessment after each ride) and three rides. The procedure thus was \emph{Baseline Assessment-Ride1-Assessment1-Ride2-Assessment2-Ride3-Assessment3}. The order of the types of music was counterbalanced across participants to avoid confounding order effects. The whole procedure in VR lasted approximately 100 mins for each participant. After the VR session, refreshments rich in electrolytes were offered. Moreover, the participants rested for 10-15 mins before leaving the premises. They were instructed to avoid driving and using heavy machinery.%
\subsection{Hypotheses and Exploratory Aims}
Based on our literature review (see \autoref{ssec:Review}), we formulated the following hypotheses: 
\begin{itemize}%
    \item \textbf{[H1]} The Calming music will have a significant mitigating effect on the intensity of the nausea-related symptoms.%
    \item \textbf{[H2]} The Joyful music will have a significant mitigating effect on overall cybersickness.%
    \item \textbf{[H3]} Cybersickness will substantially decrease verbal working memory abilities and reaction times.  
    \item \textbf{[H4]} Gaming, VR, and/or computing experience will have a mitigating effect on cybersickness.%
    \item \textbf{[H5]} Female participants will experience more intense cybersickness symptoms compared to male participants.%
\end{itemize}%
In addition to the hypotheses, based on the links between verbal working memory and reading, as well as between (positive/negative) emotions and pupil size, the current study explored the effects of cybersickness on reading ability and pupil size. Finally, this study examined the reasons for the contradictory reports in the literature on gender differences in terms of cybersickness, by examining other demographic data and their relationship or effect on cybersickness.%

\section{Results}
 \begin{figure*}[!h]
 \centering 
 \includegraphics[width= 520pt, height = 150pt]{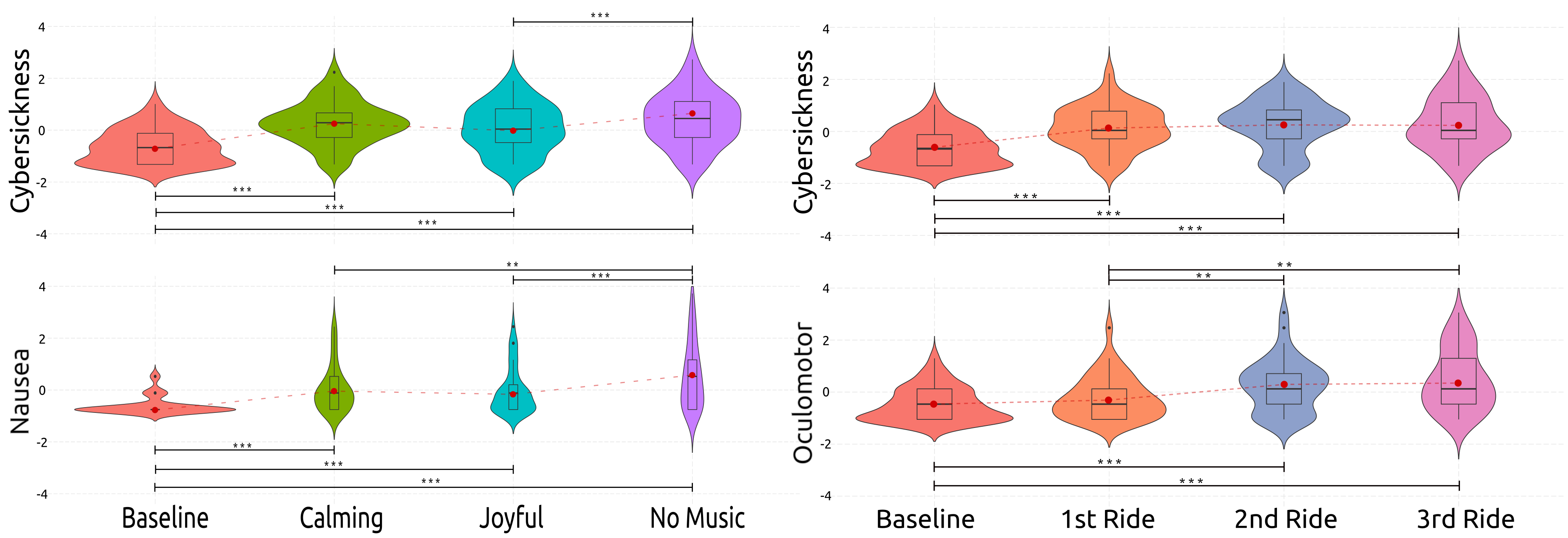}
 \caption{Overall Cybersickness (Top - Left) and Nausea (Bottom - Left) Intensities by Type of Music. Overall Cybersickness (Top - Right) and Oculomotor (Bottom - Right) Intensities by Order of Ride. *** $p<.001$, ** $p<.01$, and * $p<.05$}
 \label{fig:Rides}
 \vspace{-3mm}
\end{figure*}
All analyses were performed using R \cite{R}. As the variables violated the normality assumption, we used the $bestNormalize$ R package \cite{Peterson2020} to transform and centralize the data. The distribution of the data was then normal. Furthermore, the $afex$ (ANOVA analyses;\cite{afex}), the $ggplot2$ (plots;\cite{ggplot}), the $emmeans$ (post-hoc comparisons; \cite{emmeans}), and the $lme4$ (regression analyses;\cite{lme4}) R packages were used for performing all the analyses. Two-way repeated measures (RM) ANOVA analyses were conducted to examine the existence of interactions between the order of ride and the type of music, as well as their main effects on cybersickness. Mixed regression model analyses were performed to investigate the effect of cybersickness on cognition, motor skills, reading ability, and pupil size. Mixed model analyses were also performed to explore the effects of demographics (age, gender, education, and VR/computing/gaming experience) on cybersickness. T-tests were performed for examining differences between female and male participants. Finally, an ANCOVA was conducted, where gaming experience was a covariate, to examine whether females and males indeed experience substantially different levels of cybersickness. For the RM ANOVA, the  Greenhouse  Geisser’s correction was applied when the sphericity assumption was violated. The Bonferroni correction was applied to correct for multiple comparisons in the post hoc tests.%
\subsection{Order, Music, and Cybersickness}
The two-way ANOVAs revealed no significant interactions between music type and ride order on cybersickness. However, ride order revealed significant main effects on Nausea [$F(2.45,92.92) = 7.68$, $p <.001$, $\mathcal{\omega}^2 = .11$], Vestibular [$F(2.8,106.48) = 7.11$, $p <.001$, $\mathcal{\omega}^2 = .10$], and Oculomotor symptoms [$F(2.26,85.74) = 8.77$, $p <.001$, $\mathcal{\omega}^2 = .12$], as well as overall cybersickness [$F(2.76,105) = 11.91$, $p <.001$, $\mathcal{\omega}^2 = .15$]. Although the rides differ significantly from the baseline (see \autoref{fig:Rides}), the post hoc comparisons did not reveal significant differences among the rides for overall cybersickness, nausea, or vestibular scores. Yet, significant differences between $1st$ vs $2nd$ ride, and $1st$ vs $3rd$ ride, were observed for Oculomotor symptoms (see \autoref{fig:Rides}). However, these differences may be due to the fatigue that is under the oculomotor category of symptoms.%

In support of \textbf{H1} \& \textbf{H2}, significant main effects of music were detected for Nausea [$F(2.35,89.32) = 12.32$, $p <.001$, $\mathcal{\omega}^2 = .17$], Vestibular [$F(2.52,95.88) = 8.27$, $p <.001$, $\mathcal{\omega}^2 = .11$], and Oculomotor symptoms [$F(2.42,92.03) = 5.56$, $p <.001$, $\mathcal{\omega}^2 = .07$], as well as overall cybersickness [$F(2.53,95.97) = 14.51$, $p <.001$, $\mathcal{\omega}^2 = .17$]. However, the post hoc comparisons revealed that, except for the significant difference against the baseline, there were no other significant difference amongst the music types for vestibular and oculomotor symptoms. In contrast, regarding the Nausea symptoms, there were significant differences between Calming Music vs No Music, and Joyful Music vs No Music (see \autoref{fig:Rides}). These results are aligned with \textbf{H1} \& \textbf{H2}, since mitigating effects (substantially lower intensity of symptoms) of Calming and Joyful music are detected. Equally, in support of \textbf{H2}, there was a significant difference between Joyful Music vs No Music (see \autoref{fig:Rides}), where Joyful music substantially decreased the intensity of overall cybersickness. Finally, the ratings of joyful and calming music tracks by the participants did not show any significant correlation with cybersickness.%

\subsection{Cybersickness on Cognition, Motor Skills, Reading Ability, and Pupil Size}
\begin{figure}[!h]
 \centering 
 \includegraphics[width=\columnwidth, height = 120pt]{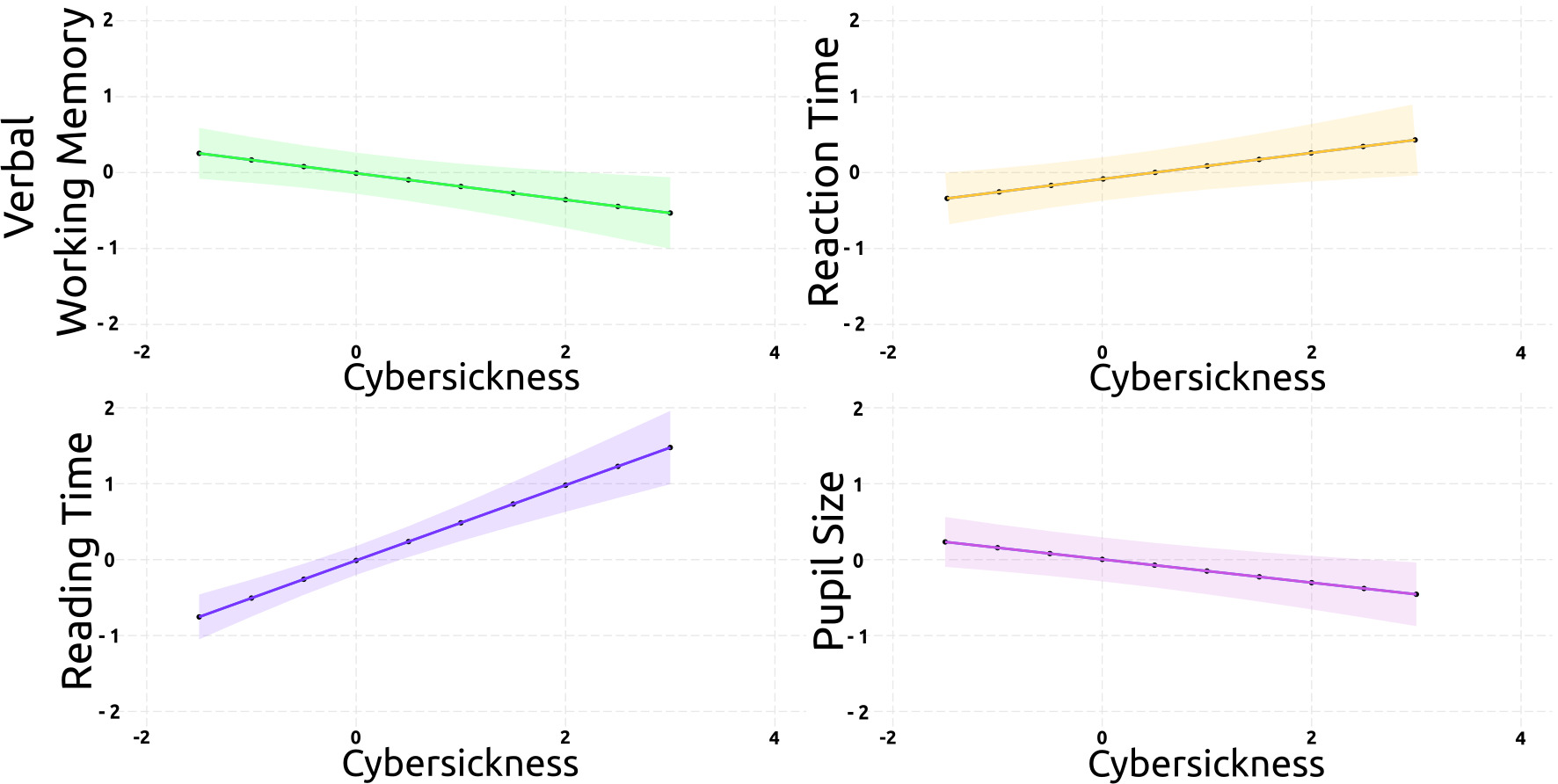}
 \caption{Overall Cybersickness's Effect on Verbal Working Memory (Top Left), Reaction Time (Top Right), Reading Time (Bottom Left), Pupil Size (Bottom Right).}
 \label{fig:CyberCog}
\end{figure}

\begin{table}[!h]
\caption{Mixed Regression Models for Cognition, Motor Skills, Reading Ability, and Pupil Size.}%
\vspace{-1.5mm}
\resizebox{\columnwidth}{1.4in}{%
\fontsize{10pt}{12pt}
\begin{tabular}{ccccccc}
\hline
Predicted &
  Predictor &
  F(1,156) &
  t(156) &
  $\mathcal{\beta}$ &
  p-value &
  $R^2$ \\ \hline
\multirow{4}{*}{\begin{tabular}[c]{@{}c@{}}Visuospatial \\ Working Memory\end{tabular}} &
  Nausea &
  1.43 &
  1.20 &
  0.10 &
  = .21 &
  .38 \\
                                & Vestibular    & 0.56  & 0.75  & 0.07  & = .45           & .37 \\
                                & Oculomotor    & 0.95  & 0.97  & 0.08  & = .33           & .38 \\
                                & Cybersickness & 0.15  & 0.39  & 0.02  & = .70           & .33 \\\hline 
\multirow{4}{*}{\begin{tabular}[c]{@{}c@{}}Verbal \\ Working Memory\end{tabular}} &
  Nausea &
  7.41 &
  -2.72 &
  -0.16 &
  \textless .01** &
  .70 \\
                                & Vestibular    & 4.27  & -2.07 & -0.16 & \textless .05*    & .69 \\
                                & Oculomotor    & 5.38  & -2.32 & -0.14 & \textless .05*    & .70 \\
                                & Cybersickness & 6.98  & -2.64 & -0.17 & \textless .01**   & .71 \\\hline 
\multirow{4}{*}{Reaction Time}  & Nausea        & 1.03  & 0.68  & 0.04  & .31           & .74 \\
                                & Vestibular    & 7.51  & 1.79  & 0.13  & \textless .01**   & .74 \\
                                & Oculomotor    & 0.75  & 0.27  & 0.02  & = .39           & .74 \\
                                & Cybersickness & 3.84  & 0.97  & 0.09  & \textless .05*    & .74 \\\hline 
\multirow{4}{*}{Attention Time} & Nausea        & 2.60  & 1.61  & 0.13  & = .11           & .25 \\
                                & Vestibular    & 3.10  & 1.76  & 0.18  & = .06           & .27 \\
                                & Oculomotor    & 3.05  & 1.75  & 0.15  & = .06           & .26 \\
                                & Cybersickness & 2.24  & 1.50  & 0.14  & = .11           & .26 \\\hline 
\multirow{4}{*}{Motor Time}     & Nausea        & 0.22  & 0.47  & 0.03  & = .47           & .60 \\
                                & Vestibular    & 0.05  & 0.21  & 0.02  & = .26           & .60 \\
                                & Oculomotor    & 1.66  & 1.29  & 0.09  & = .20           & .60 \\
                                & Cybersickness & 0.23  & 0.47  & 0.04  & = .65           & .60 \\\hline 
\multirow{4}{*}{Reading Time} &
  Nausea &
  38.63 &
  6.22 &
  0.43 &
  \textless .001*** &
  .49 \\
                                & Vestibular    & 32.13 & 5.67  & 0.50  & \textless .001*** & .44 \\
                                & Oculomotor    & 11.3  & 3.36  & 0.26  & \textless .001*** & .44 \\
                                & Cybersickness & 24.39 & 4.94 & 1.51  & \textless .001*** & .61 \\\hline 
\multirow{4}{*}{Pupil Size}   & Nausea        & 2.59  & -1.61 & -0.07 & = .11           & .82 \\
                                & Vestibular    & 4.95  & -2.22 & -0.13 & \textless .05*    & .83 \\
                                & Oculomotor    & 10.5  & -3.24 & -0.15 & \textless .001*** & .83 \\
                                & Cybersickness & 8.69  & -2.95 & -0.15 & \textless .001*** & .83 \\ \hline 
\end{tabular}%
}
\label{tab:CogModels}%
 \vspace{-6mm}
\end{table}
Cybersickness had an effect on verbal working memory, albeit there was no effect on visuospatial working memory. In line with \textbf{H3}, overall cybersickness had a negative effect on verbal working memory (see \autoref{tab:CogModels} and \autoref{fig:CyberCog}). While all types of cybersickness symptoms had an effect on verbal working memory, the nausea- and vestibular-related symptoms had a slightly higher negative effect (see t value and $\mathcal{\beta}$ coefficient in \autoref{tab:CogModels}). Furthermore, in support of \textbf{H3}, overall cybersickness had an effect on reaction time. As displayed in \autoref{fig:CyberCog}, higher cybersickness increases the reaction time, postulating a substantial deceleration.%

However, only the vestibular symptoms appeared to have an effect on reaction time, while there were no significant effects on reaction time by nausea and oculomotor symptoms (see \autoref{tab:CogModels}).Furthermore, cybersickness (overall and symptoms) had no effect on attention time or motor time. Interestingly, the effects on motor time were very small (see $\mathcal{\beta}$ in \autoref{tab:CogModels}). In contrast, the effects on attention time tended to be significant, and their size was large (see $\mathcal{\beta}$ in \autoref{tab:CogModels}). Given that reaction time is the addition of attention and motor time, the results indicate that the effects on reaction time predominantly stem from a substantial deceleration of attentional processing speed. Moreover, the exploratory analyses revealed significant effects  of cybersickness on reading time and pupil size (see \autoref{fig:CyberCog}). Every symptom type substantially prolonged reading time (see \autoref{tab:CogModels}). Lastly, except for nausea symptoms, the other types of cybersickness symptoms substantially decreased pupil size.%

\subsection{Cybersickness and Individual Differences}
\begin{figure}[!h]
 \centering 
 \includegraphics[width=\columnwidth, height = 130pt]{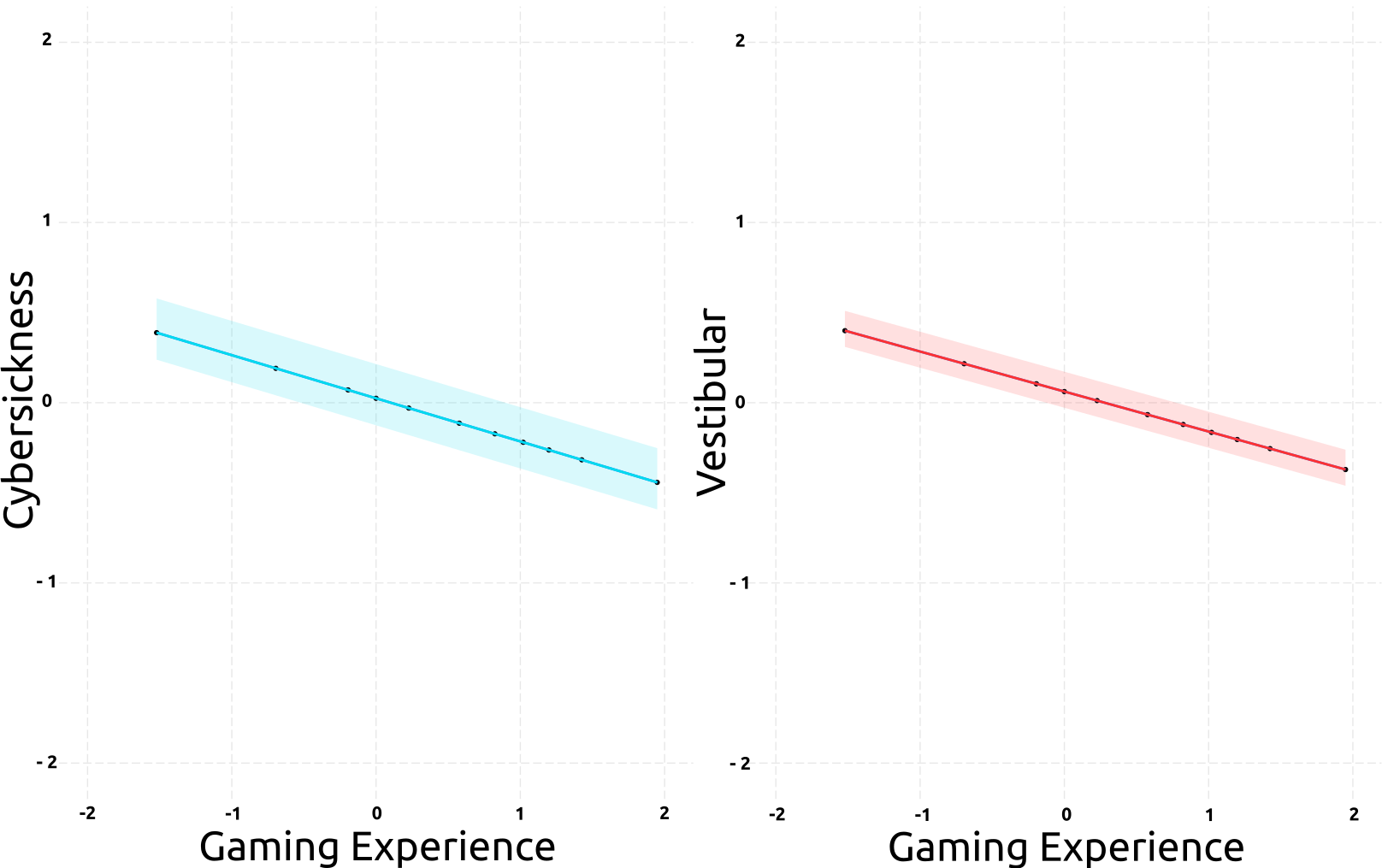}
 \caption{Gaming Experience Effect on Overall Cybersickness (Left) Vestibular (Right) Intensity.}
 \label{fig:GameCyber}
 \vspace{-3mm}
\end{figure}

As displayed in \autoref{tab:Demos}, gender and gaming experience were the only demographics that predicted cybersickness intensity. In line with \textbf{H4}, individuals with higher gaming experience had substantially lower overall cybersickness, as well as vestibular symptoms (see \autoref{fig:GameCyber}). Nonetheless, gaming experience did not significantly predict nausea or oculomotor symptoms. On the other hand, the user's gender significantly predicted the intensity of cybersickness. In support of \textbf{H5}, female participants experienced substantially more intense cybersickness symptoms compared to male participants [$t(37)=3.08$, $p=.004$].%
 \begin{table}[!h]
 \vspace{-2mm}
\caption{Mixed Regression Models for Cybersickness: Demographics. \\ MSA \& MSB = motion sickness susceptibility indexes}%
\resizebox{\columnwidth}{0.8in}{%
\fontsize{10pt}{10pt}
\begin{tabular}{cccccc}
\hline
Predictor                                                      & F(1,156) & t(156) & $\mathcal{\beta}$ coefficient & p-value         & $R^2$ \\ \hline
Age         & 0.05 & -0.22 & -0.03 & = .83           & .46 \\
Education   & 0.03 & 0.16  & 0.02  & = .87           & .46 \\
Gender      & 9.47 & -3.08 & -0.65 & \textless .001*** & .46 \\
\begin{tabular}[c]{@{}c@{}}Computing\\ Experience\end{tabular} & 0.25     & 0.50   & 0.07                          & = .62         & .46   \\
\begin{tabular}[c]{@{}c@{}}VR\\ Experience\end{tabular}        & 1.00     & -1.00  & -0.12                         & = .32         & .46   \\
\begin{tabular}[c]{@{}c@{}}Gaming\\ Experience\end{tabular}    & 4.00     & -2.00  & -0.24                         & \textless .01** & .46   \\
MSA - Child & 2.06 & 1.43  & 0.10  & = .15           & .42 \\
MSB - Adult & 0.88 & 0.94  & 0.06  & = .35           & .42 \\\hline
\end{tabular}%
}
\label{tab:Demos}%
 \vspace{-5mm}
\end{table}

\begin{figure}[!h]
  \vspace{-2mm}
 \centering 
 \includegraphics[width= \columnwidth]{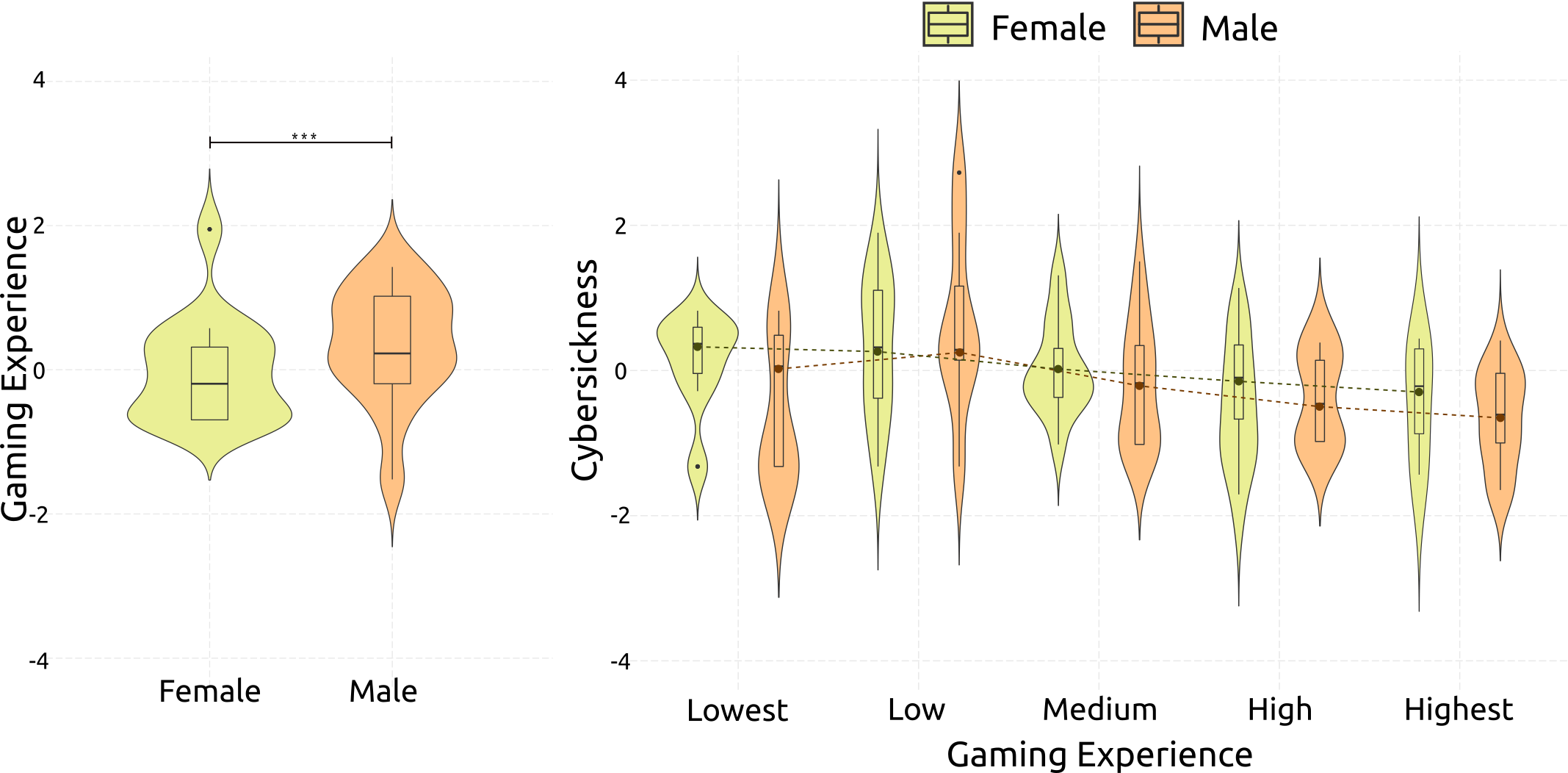}
 \caption{Gaming Experience Difference between Female and Male Participants (Left). Cybersickness in Female and Male Participants per Gaming Experience Level (Right).}
 \label{fig:Gender}
 \vspace{-6mm}
\end{figure}

 However, another significant difference was identified between female and male participants. As illustrated in \autoref{fig:Gender}, female participants had a significantly lower gaming experience than the male participants [$t(37)=-2.76$, $p<.001$]. Given that gaming experience was a significant predictor of cybersickness with a large $\mathcal{\beta}$ (see \autoref{tab:Demos}),gaming experience was considered as a covariate in the ANCOVA for assessing gender's effect on cybersickness. In contrast with \textbf{H5}, there was not a significant main effect of gender on cybersickness [$F(1,30) = 0.93$, $p=.45$, $\mathcal{\omega}^2 = .01e-03$], postulating that the observed differences in cybersickness were attributed to the difference in gaming experience between female and male participants. Finally, post hoc comparisons of the marginal means indicated that there were not any significant differences between female and male users when gaming experience was matched between them (see \autoref{fig:Gender}).%
\section{Discussion}
\subsection{Music and Cybersickness}
Calming music was found to have a significant mitigating effect on nausea symptoms induced by cybersickness. These results are in line with the findings in clinical populations (e.g., cancer patients), where calming/relaxing music reduced nausea induced by trauma (e.g., brain injury) or drugs (e.g., chemotherapy) \cite{Karagozoglu2013}. However, pleasant/joyful music was also successful in substantially alleviating nausea. Notably, only the pleasant/joyful music was able to substantially mitigate the intensity of overall cybersickness. This outcome is in line with the findings of previous studies on motion sickness (e.g., \cite{Peck2020,Keshavarz2014}). Nevertheless, in contrast with the study of Peck et al. \cite{Peck2020}, the ratings of joyful and/or calming music did not correlate with cybersickness, postulating that the liking of a song did not play a significant role. We infer that the reason for this disagreement is that the music tracks used in our study were derived from a  relatively large scale online study. Consequently, and as observed in the ratings of the VR study, both tracks were highly rated by the participants.%

The mitigation of nausea symptoms by music has been suggested to arise from a synchronization of the autonomous nervous system \cite{Bernardi}. However, this theory fails to explain how pleasant/joyful music is even more effective in mitigating overall motion- or cybersickness. An alternative explanation is that pleasant music has a combined effect, where it distracts the user from stimuli that induce symptoms, and also elicits an arousal of positive emotions \cite{Peck2020}. Our results suggest that joyful music was more successful than calming music, since it mitigated both nausea and overall cybersickness. Also, the ratings indicated that the song was perceived as highly joyful by the vast majority of the participants.%

Moreover, pupil size has been previously seen as a biomarker of affective state, where bigger size indicates a positive and smaller size indicates a negative affective state  \cite{Partala}. Our results demonstrated a comparable pattern in cybersickness, where more intense cybersickness induced a smaller diameter (i.e., more negative affective state), and less intense a bigger diameter of the pupils (i.e., more positive affective state). Notably, the mixed regression model of cybersickness explained 83\% of the variance of pupil size. Thus, these results further support the notion that pleasant music is efficient in mitigating cybersickness due to  distraction from stimuli, as well as an elicitation of positive emotions. In summary, the implementation of music appears to be a cost-effective, suitable (i.e., without preventing other interactions or breaking the immersion), and efficient technique for mitigating cybersickness in immersive VR.%

\subsection{Cybersickness's Effects on Cognitive and Motor Skills}
In the current study, cybersickness was found to substantially affect performance on a verbal working memory task. This outcome aligns with the previous observation in motion-sickness, where verbal working memory was significantly decreased \cite{Dahlman2009}. However, in our study, the visuospatial working memory remained intact regardless of cybersickness. These results agree with the previous findings where visuospatial working memory was not affected by cybersickness \cite{Mittelstaedt2019}. Taken together, our and previous results postulate that cybersickness negatively affects only the verbal working memory, while the non-verbal processes remain intact. This could indicate that cybersickness may affect also higher cognitive functions, which require an intact verbal working memory, such as episodic memory and learning (see \cite{Cowan} for the link between these cognitive functions).%

Moreover, in our study, reading time was substantially increased during cybersickness, connoting slower reading speed. Considering the association between verbal working memory and reading ability \cite{Cowan}, this result further suggests that the verbal processes are being affected by cybersickness. Nevertheless, regarding reading ability, this observation in our study could have stemmed from a confusion in decision making (e.g., selecting an appropriate response to the question, which motivated a re-reading of the question) rather than decreased reading speed. Thus, further research is required for understanding whether reading ability is indeed affected, or there is another cognitive process (e.g., decision making or comprehension) that is affected by cybersickness.%

Furthermore, in line with previous findings \cite{Mittelstaedt2019,Nalivaiko2015,Nesbitt2017}, a substantial deceleration of reaction time was detected. However, we did not find a significant deceleration of the components of psychomotor skills. Neither attentional processing time, nor motor time was significantly affected. Nevertheless, the effect of cybersickness on attentional processing speed tended to be significant, which supports the findings of Mittelstaedt et al. \cite{Mittelstaedt2019}, where visual attention speed was significantly slowed by cybersickness. In our study, the effect of cybersickness on attentional speed was large, while the effect size on motor speed was small. Considering that reaction speed consists of attentional and motor speed, these results indicate that the significant decrease in reaction speed is attributed predominantly to the decrease of the attentional processing speed. These effects of cybersickness are crucial in VR applications in education and/or occupational training, where intact attention and reflexes are required.%
\subsection{Gender and Gaming Experience in Cybersickness}
Only gender and gaming experience of the user were found to significantly predict the intensity of the experienced cybersickness symptoms. In contrast with research on motion sickness, where MSSQ is efficient in predicting the intensities of motion sickness \cite{GOLDING}, the scores of MSSQ were not efficient in predicting the intensity of cybersickness.This inability of MSSQ to predict cybersickness agrees with the fact that cybersickness substantially differs from simulator and motion sickness in terms of stimuli eliciting the symptoms (e.g., vection instead of motion) and the frequency of symptoms (e.g., disorientation is the most frequent in cybersickness) \cite{Stanney1997,Davis2014}. Furthermore, age did not appear to predict cybersickness. However, it should be considered that the participants in the current study were young adults. The effect of age should be examined in an age-diverse population. Lastly, neither VR nor computing experience predicted cybersickness. Regarding the effect of VR experience on cybersickness, a systematic review showed there are discrepant results in the literature, where some studies reported a resilience in experienced VR users, while others did not detect differences \cite{Tian2022}. Though, previous studies suffered from did not meticulously measuring VR experience \cite{Tian2022}. In this study, VR experience, as well as computing and gaming experience, were measured in a standardized and effective way \cite{Kourtesis2019, Smith}, yet, VR experience failed to predict cybersickness. However, the population did not appear to have high levels of experience. Hence, the effect of VR experience should be further studied in a population with greater variability in terms of VR experience.%

In line with the findings of \cite{Stanney2020,Chatta}, we did find significant differences between male and female participants in terms of cybersickness intensity. However, Stanney et al. \cite{Stanney2020} suggested that these differences derive from the IPD of the VR headset. Indeed, in their $2nd$ experiment, they did not find any difference when participants found the IPD comfortable. However, in our study, we used an HTC Vive Pro Eye, which is comfortable for everyone. Also, every participant went through eye-tracking calibration, where the IPD is adjusted for the successful completion of calibration. Hence, every participant had an appropriate IPD and felt comfortable. Regardless of the IPD adjustment and comfort, differences in cybersickness intensities were present. Thus, other factors appear to modulate the differences between female and male users.%

In our study, gaming experience was a significant predictor of cybersickness, which aligns with the findings of the large scale study of Weech et al. \cite{Weech2020}. Also, our findings on cybersickness agree with previous studies, where higher gaming experience associated with augmented resilience to simulator sickness \cite{Himmels} and cybersickness \cite{Grassini}. Moreover, the gaming experience of female participants was significantly smaller compared to male participants. Previous studies who reported higher cybersickness in females compared to males (e.g., \cite{Stanney2020,Chatta}) did not consider gaming experience for explaining these differences. In our study, when we controlled for gaming experience, the intensities of cybersickness did not significantly differ between female and male users. This result agrees with the outcome of a meta-analysis of studies on cybersickness, which suggested no difference between males and females \cite{Saredakis2020}. Importantly, it postulates that gaming experience is a factor that substantially modulates the differences between female and male VR users. Higher gaming experience indicated a resilience to cybersickness, while lower gaming experience indicates a susceptibility to it. However,it should be noted that this study considered experience as a combination of the ability (i.e., how capable someone is of using VR/PC or playing games) and the frequency (i.e., how often someone uses VR/PC or plays games) pertinent to VR, computers, and video games. Gaming experience  should thus be efficiently evaluated and considered in future studies on cybersickness.%
\subsection{Limitations and Future Studies}
The current study has several limitations that should be considered. Firstly, the population consisted of young adults. The effects of cybersickness should be examined in a more age-diverse population, where the role of individual differences and demographics could be further studied. In addition, cybersickness intensities in the current study mainly ranged between mild to moderate symptoms. This probably is a result of the parsimonious inclusion criterion that we used for the MSSQ scores (i.e., excluding individuals who scored higher than the 75th percentile who could experience substantially stronger symptoms). Since cybersickness intensities appear to substantially vary across individuals, future studies should examine the effects of music on cybersickness and cybersickness effects on cognitive and motor skills in a population that may experience stronger symptoms. Also, the VR exposure in this study lasted approximately 70 minutes (100 mins for the whole protocol, pre and post exposure) per participant, which is considered a long exposure to VR. It would be of interest to replicate the findings of the present study in a shorter exposure. Lastly, the current study used only working memory tests and a classical psychomotor task. Future studies should attempt to examine other cognitive functions (e.g., episodic memory or decision making) and motor skills (e.g., tasks that require fine motor skills and accuracy).%
\section{Conclusion}
Music, especially joyful, appeared a cost-effective and efficient method for mitigating nausea and overall cybersickness. Also, cybersickness was found to affect only verbal working memory but not non-verbal (visuospatial) working memory. Given the important role of verbal working memory in higher cognitive functions, this effect of cybersickness should be considered in VR applications for educational, training, and clinical purposes. Moreover, Pupil size was substantially affected by cybersickness, and it can thus be considered as a biomarker of cybersickness. Higher gaming experience was associated with lower cybersickness, and explained the differences between female and male users in terms of cybersickness. Gaming experience should be appropriately evaluated and considered in future studies on cybersickness.%

\acknowledgments{%
The authors would like to thank the uCreate Studio of the University of Edinburgh for providing them with VR equipment and tech support.%
}

\bibliographystyle{abbrv-doi-hyperref}

\bibliography{my}

\begin{thebibliography}{10}

\bibitem{lme4}
D.~Bates, M.~M{\"a}chler, B.~Bolker, and S.~Walker.
\newblock Fitting linear mixed-effects models using {lme4}.
\newblock {\em Journal of Statistical Software}, 67(1):1--48, 2015.
  \href{https://doi.org/10.18637/jss.v067.i01}
{doi: {{%
10\hspace{.1pt}\discretionary{.}{%
}{.}\hspace{.4pt}18637\discretionary{/}{%
}{/}jss\hspace{.1pt}\discretionary{.}{%
}{.}\hspace{.4pt}v067\hspace{.1pt}\discretionary{.}{%
}{.}\hspace{.4pt}i01}}}


\bibitem{bmlTUX}
A.~O. Bebko and N.~F. Troje.
\newblock bmltux: Design and control of experiments in virtual reality and
  beyond, July 2020. \href{https://doi.org/10.1177/2041669520938400}
{doi: {{%
10\hspace{.1pt}\discretionary{.}{%
}{.}\hspace{.4pt}1177\discretionary{/}{%
}{/}2041669520938400}}}


\bibitem{Bernardi}
N.~F. Bernardi, E.~Codrons, R.~di~Leo, M.~Vandoni, F.~Cavallaro, G.~Vita, and
  L.~Bernardi.
\newblock Increase in synchronization of autonomic rhythms between individuals
  when listening to music.
\newblock {\em Frontiers in Physiology}, 8, 2017.
  \href{https://doi.org/10.3389/fphys.2017.00785}
{doi: {{%
10\hspace{.1pt}\discretionary{.}{%
}{.}\hspace{.4pt}3389\discretionary{/}{%
}{/}fphys\hspace{.1pt}\discretionary{.}{%
}{.}\hspace{.4pt}2017\hspace{.1pt}\discretionary{.}{%
}{.}\hspace{.4pt}00785}}}


\bibitem{Chatta}
U.~A. Chattha, U.~I. Janjua, F.~Anwar, T.~M. Madni, M.~F. Cheema, and S.~I.
  Janjua.
\newblock Motion sickness in virtual reality: An empirical evaluation.
\newblock {\em IEEE Access}, 8:130486--130499, 2020.
  \href{https://doi.org/10.1109/ACCESS.2020.3007076}
{doi: {{%
10\hspace{.1pt}\discretionary{.}{%
}{.}\hspace{.4pt}1109\discretionary{/}{%
}{/}ACCESS\hspace{.1pt}\discretionary{.}{%
}{.}\hspace{.4pt}2020\hspace{.1pt}\discretionary{.}{%
}{.}\hspace{.4pt}3007076}}}


\bibitem{Cheung2005}
B.~Cheung and K.~Hofer.
\newblock Desensitization to strong vestibular stimuli improves tolerance to
  simulated aircraft motion.
\newblock {\em Aviation, Space, and Environmental Medicine}, 76(12):1099--1104,
  2005.

\bibitem{Corsi}
P.~M. Corsi.
\newblock Human memory and the medial temporal region of the brain.
\newblock 1972.

\bibitem{Cowan}
N.~Cowan.
\newblock Working memory underpins cognitive development, learning, and
  education.
\newblock {\em Educational Psychology Review}, 26(2):197--223, 2014.
  \href{https://doi.org/10.1007/s10648-013-9246-y}
{doi: {{%
10\hspace{.1pt}\discretionary{.}{%
}{.}\hspace{.4pt}1007\discretionary{/}{%
}{/}s10648\discretionary{%
}{-}{-}013\discretionary{%
}{-}{-}9246\discretionary{%
}{-}{-}y}}}


\bibitem{Cruz-Neira2018}
C.~Cruz-Neira, M.~Fernández, and C.~Portalés.
\newblock Virtual reality and games.
\newblock {\em Multimodal Technologies and Interaction}, 2(1), 2018.
  \href{https://doi.org/10.3390/mti2010008}
{doi: {{%
10\hspace{.1pt}\discretionary{.}{%
}{.}\hspace{.4pt}3390\discretionary{/}{%
}{/}mti2010008}}}


\bibitem{Dahlman2009}
J.~Dahlman, A.~Sjörs, J.~Lindström, T.~Ledin, and T.~Falkmer.
\newblock Performance and autonomic responses during motion sickness.
\newblock {\em Human Factors}, 51(1):56--66, 2009.
\newblock PMID: 19634309. \href{https://doi.org/10.1177/0018720809332848}
{doi: {{%
10\hspace{.1pt}\discretionary{.}{%
}{.}\hspace{.4pt}1177\discretionary{/}{%
}{/}0018720809332848}}}


\bibitem{Davis2014}
S.~Davis, K.~Nesbitt, and E.~Nalivaiko.
\newblock A systematic review of cybersickness.
\newblock In {\em Proceedings of the 2014 Conference on Interactive
  Entertainment}, IE2014, p. 1–9. Association for Computing Machinery, New
  York, NY, USA, 2014. \href{https://doi.org/10.1145/2677758.2677780}
{doi: {{%
10\hspace{.1pt}\discretionary{.}{%
}{.}\hspace{.4pt}1145\discretionary{/}{%
}{/}2677758\hspace{.1pt}\discretionary{.}{%
}{.}\hspace{.4pt}2677780}}}


\bibitem{Deary}
I.~J. Deary, D.~Liewald, and J.~Nissan.
\newblock A free, easy-to-use, computer-based simple and four-choice reaction
  time programme: the deary-liewald reaction time task.
\newblock {\em Behavior research methods}, 43(1):258—268, March 2011.
  \href{https://doi.org/10.3758/s13428-010-0024-1}
{doi: {{%
10\hspace{.1pt}\discretionary{.}{%
}{.}\hspace{.4pt}3758\discretionary{/}{%
}{/}s13428\discretionary{%
}{-}{-}010\discretionary{%
}{-}{-}0024\discretionary{%
}{-}{-}1}}}


\bibitem{Dennison2016}
M.~S. Dennison, A.~Z. Wisti, and M.~D’Zmura.
\newblock Use of physiological signals to predict cybersickness.
\newblock {\em Displays}, 44:42--52, 2016.
\newblock Contains Special Issue Articles – Proceedings of the 4th Symposium
  on Liquid Crystal Photonics (SLCP 2015).
  \href{https://doi.org/10.1016/j.displa.2016.07.002}
{doi: {{%
10\hspace{.1pt}\discretionary{.}{%
}{.}\hspace{.4pt}1016\discretionary{/}{%
}{/}j\hspace{.1pt}\discretionary{.}{%
}{.}\hspace{.4pt}displa\hspace{.1pt}\discretionary{.}{%
}{.}\hspace{.4pt}2016\hspace{.1pt}\discretionary{.}{%
}{.}\hspace{.4pt}07\hspace{.1pt}\discretionary{.}{%
}{.}\hspace{.4pt}002}}}


\bibitem{Dixon}
P.~Dixon, J.-A. LeFevre, and L.~C. Twilley.
\newblock Word knowledge and working memory as predictors of reading skill.
\newblock {\em Journal of Educational Psychology}, 80(4):465, 1988.
  \href{https://doi.org/10.1037/0022-0663.80.4.465}
{doi: {{%
10\hspace{.1pt}\discretionary{.}{%
}{.}\hspace{.4pt}1037\discretionary{/}{%
}{/}0022\discretionary{%
}{-}{-}0663\hspace{.1pt}\discretionary{.}{%
}{.}\hspace{.4pt}80\hspace{.1pt}\discretionary{.}{%
}{.}\hspace{.4pt}4\hspace{.1pt}\discretionary{.}{%
}{.}\hspace{.4pt}465}}}


\bibitem{Emmelkamp2021}
P.~M. Emmelkamp and K.~Meyerbr\"{o}ker.
\newblock Virtual reality therapy in mental health.
\newblock {\em Annual Review of Clinical Psychology}, 17(1):495--519, 2021.
\newblock PMID: 33606946.
  \href{https://doi.org/10.1146/annurev-clinpsy-081219-115923}
{doi: {{%
10\hspace{.1pt}\discretionary{.}{%
}{.}\hspace{.4pt}1146\discretionary{/}{%
}{/}annurev\discretionary{%
}{-}{-}clinpsy\discretionary{%
}{-}{-}081219\discretionary{%
}{-}{-}115923}}}


\bibitem{Farmani2020}
Y.~Farmani and R.~J. Teather.
\newblock Evaluating discrete viewpoint control to reduce cybersickness in
  virtual reality.
\newblock {\em Virtual Reality}, 24(4):645–664, dec 2020.
  \href{https://doi.org/10.1007/s10055-020-00425-x}
{doi: {{%
10\hspace{.1pt}\discretionary{.}{%
}{.}\hspace{.4pt}1007\discretionary{/}{%
}{/}s10055\discretionary{%
}{-}{-}020\discretionary{%
}{-}{-}00425\discretionary{%
}{-}{-}x}}}


\bibitem{Freedman}
D.~Freedman.
\newblock From association to causation via regression.
\newblock {\em Advances in Applied Mathematics}, 18(1):59--110, 1997.
  \href{https://doi.org/10.1006/aama.1996.0501}
{doi: {{%
10\hspace{.1pt}\discretionary{.}{%
}{.}\hspace{.4pt}1006\discretionary{/}{%
}{/}aama\hspace{.1pt}\discretionary{.}{%
}{.}\hspace{.4pt}1996\hspace{.1pt}\discretionary{.}{%
}{.}\hspace{.4pt}0501}}}


\bibitem{GOLDING}
J.~F. Golding.
\newblock Predicting individual differences in motion sickness susceptibility
  by questionnaire.
\newblock {\em Personality and Individual Differences}, 41(2):237--248, 2006.
  \href{https://doi.org/10.1016/j.paid.2006.01.012}
{doi: {{%
10\hspace{.1pt}\discretionary{.}{%
}{.}\hspace{.4pt}1016\discretionary{/}{%
}{/}j\hspace{.1pt}\discretionary{.}{%
}{.}\hspace{.4pt}paid\hspace{.1pt}\discretionary{.}{%
}{.}\hspace{.4pt}2006\hspace{.1pt}\discretionary{.}{%
}{.}\hspace{.4pt}01\hspace{.1pt}\discretionary{.}{%
}{.}\hspace{.4pt}012}}}


\bibitem{Golding2015}
J.~F. Golding and M.~A. Gresty.
\newblock Pathophysiology and treatment of motion sickness.
\newblock {\em Current opinion in neurology}, 28(1):83--88, 2015.
  \href{https://doi.org/10.1097/WCO.0000000000000163}
{doi: {{%
10\hspace{.1pt}\discretionary{.}{%
}{.}\hspace{.4pt}1097\discretionary{/}{%
}{/}WCO\hspace{.1pt}\discretionary{.}{%
}{.}\hspace{.4pt}0000000000000163}}}


\bibitem{Graff796}
V.~Graff, L.~Cai, I.~Badiola, and N.~M. Elkassabany.
\newblock Music versus midazolam during preoperative nerve block placements: a
  prospective randomized controlled study.
\newblock {\em Regional Anesthesia \& Pain Medicine}, 44(8):796--799, 2019.
  \href{https://doi.org/10.1136/rapm-2018-100251}
{doi: {{%
10\hspace{.1pt}\discretionary{.}{%
}{.}\hspace{.4pt}1136\discretionary{/}{%
}{/}rapm\discretionary{%
}{-}{-}2018\discretionary{%
}{-}{-}100251}}}


\bibitem{Grassini}
S.~Grassini, K.~Laumann, and A.~K. Luzi.
\newblock Association of individual factors with simulator sickness and sense
  of presence in virtual reality mediated by head-mounted displays (hmds).
\newblock {\em Multimodal Technologies and Interaction}, 5(3), 2021.
  \href{https://doi.org/10.3390/mti5030007}
{doi: {{%
10\hspace{.1pt}\discretionary{.}{%
}{.}\hspace{.4pt}3390\discretionary{/}{%
}{/}mti5030007}}}


\bibitem{Himmels}
C.~Himmels, A.~R. Czupala, T.~Fahm\"{u}ller, B.~Fuchs, L.~Hauf, L.~Heidler, and
  A.~Riener.
\newblock The influence of gaming experience and optic flow on simulator
  sickness: Insights from a driving simulator study.
\newblock In {\em Adjunct Proceedings of the 14th International Conference on
  Automotive User Interfaces and Interactive Vehicular Applications},
  AutomotiveUI '22, p. 149–152. Association for Computing Machinery, New
  York, NY, USA, 2022. \href{https://doi.org/10.1145/3544999.3552532}
{doi: {{%
10\hspace{.1pt}\discretionary{.}{%
}{.}\hspace{.4pt}1145\discretionary{/}{%
}{/}3544999\hspace{.1pt}\discretionary{.}{%
}{.}\hspace{.4pt}3552532}}}


\bibitem{Karagozoglu2013}
S.~Karagozoglu, F.~Tekyasar, and F.~A. Yilmaz.
\newblock Effects of music therapy and guided visual imagery on
  chemotherapy-induced anxiety and nausea-vomiting.
\newblock {\em Journal of Clinical Nursing}, 22(1-2):39—50, January 2013.
  \href{https://doi.org/10.1111/jocn.12030}
{doi: {{%
10\hspace{.1pt}\discretionary{.}{%
}{.}\hspace{.4pt}1111\discretionary{/}{%
}{/}jocn\hspace{.1pt}\discretionary{.}{%
}{.}\hspace{.4pt}12030}}}


\bibitem{Keshavarz2014}
B.~Keshavarz and H.~Hecht.
\newblock Pleasant music as a countermeasure against visually induced motion
  sickness.
\newblock {\em Applied Ergonomics}, 45(3):521--527, 2014.
  \href{https://doi.org/10.1016/j.apergo.2013.07.009}
{doi: {{%
10\hspace{.1pt}\discretionary{.}{%
}{.}\hspace{.4pt}1016\discretionary{/}{%
}{/}j\hspace{.1pt}\discretionary{.}{%
}{.}\hspace{.4pt}apergo\hspace{.1pt}\discretionary{.}{%
}{.}\hspace{.4pt}2013\hspace{.1pt}\discretionary{.}{%
}{.}\hspace{.4pt}07\hspace{.1pt}\discretionary{.}{%
}{.}\hspace{.4pt}009}}}


\bibitem{Kim2021}
J.~Kim, S.~Palmisano, W.~Luu, and S.~Iwasaki.
\newblock Effects of linear visual-vestibular conflict on presence, perceived
  scene stability and cybersickness in the oculus go and oculus quest.
\newblock {\em Frontiers in Virtual Reality}, 2, 2021.
  \href{https://doi.org/10.3389/frvir.2021.582156}
{doi: {{%
10\hspace{.1pt}\discretionary{.}{%
}{.}\hspace{.4pt}3389\discretionary{/}{%
}{/}frvir\hspace{.1pt}\discretionary{.}{%
}{.}\hspace{.4pt}2021\hspace{.1pt}\discretionary{.}{%
}{.}\hspace{.4pt}582156}}}


\bibitem{Kourtesis2019}
P.~Kourtesis, S.~Collina, L.~A.~A. Doumas, and S.~E. MacPherson.
\newblock Validation of the virtual reality neuroscience questionnaire: Maximum
  duration of immersive virtual reality sessions without the presence of
  pertinent adverse symptomatology.
\newblock {\em Frontiers in Human Neuroscience}, 13, 2019.
  \href{https://doi.org/10.3389/fnhum.2019.00417}
{doi: {{%
10\hspace{.1pt}\discretionary{.}{%
}{.}\hspace{.4pt}3389\discretionary{/}{%
}{/}fnhum\hspace{.1pt}\discretionary{.}{%
}{.}\hspace{.4pt}2019\hspace{.1pt}\discretionary{.}{%
}{.}\hspace{.4pt}00417}}}


\bibitem{Kourtesis2019b}
P.~Kourtesis, D.~Korre, S.~Collina, L.~A.~A. Doumas, and S.~E. MacPherson.
\newblock Guidelines for the development of immersive virtual reality software
  for cognitive neuroscience and neuropsychology: The development of virtual
  reality everyday assessment lab (vr-eal), a neuropsychological test battery
  in immersive virtual reality.
\newblock {\em Frontiers in Computer Science}, 1, 2020.
  \href{https://doi.org/10.3389/fcomp.2019.00012}
{doi: {{%
10\hspace{.1pt}\discretionary{.}{%
}{.}\hspace{.4pt}3389\discretionary{/}{%
}{/}fcomp\hspace{.1pt}\discretionary{.}{%
}{.}\hspace{.4pt}2019\hspace{.1pt}\discretionary{.}{%
}{.}\hspace{.4pt}00012}}}


\bibitem{KourtesisCSQ}
P.~Kourtesis, J.~Linnell, R.~Amir, F.~Argelaguet, and S.~E. MacPherson.
\newblock Cybersickness in virtual reality questionnaire (csq-vr): A validation
  and comparison against ssq and vrsq.
\newblock {\em Virtual Worlds}, 2(1):16--35, 2023.
  \href{https://doi.org/10.3390/virtualworlds2010002}
{doi: {{%
10\hspace{.1pt}\discretionary{.}{%
}{.}\hspace{.4pt}3390\discretionary{/}{%
}{/}virtualworlds2010002}}}


\bibitem{Kourtesis2021}
P.~Kourtesis and S.~E. MacPherson.
\newblock How immersive virtual reality methods may meet the criteria of the
  national academy of neuropsychology and american academy of clinical
  neuropsychology: A software review of the virtual reality everyday assessment
  lab (vr-eal).
\newblock {\em Computers in Human Behavior Reports}, 4:100151, 2021.
  \href{https://doi.org/10.1016/j.chbr.2021.100151}
{doi: {{%
10\hspace{.1pt}\discretionary{.}{%
}{.}\hspace{.4pt}1016\discretionary{/}{%
}{/}j\hspace{.1pt}\discretionary{.}{%
}{.}\hspace{.4pt}chbr\hspace{.1pt}\discretionary{.}{%
}{.}\hspace{.4pt}2021\hspace{.1pt}\discretionary{.}{%
}{.}\hspace{.4pt}100151}}}


\bibitem{LaViola2000}
J.~J. LaViola.
\newblock A discussion of cybersickness in virtual environments.
\newblock {\em SIGCHI Bull.}, 32(1):47–56, jan 2000.
  \href{https://doi.org/10.1145/333329.333344}
{doi: {{%
10\hspace{.1pt}\discretionary{.}{%
}{.}\hspace{.4pt}1145\discretionary{/}{%
}{/}333329\hspace{.1pt}\discretionary{.}{%
}{.}\hspace{.4pt}333344}}}


\bibitem{emmeans}
R.~V. Lenth.
\newblock {\em emmeans: Estimated Marginal Means, aka Least-Squares Means},
  2022.
\newblock R package version 1.7.5.

\bibitem{Melo2018}
M.~Melo, J.~Vasconcelos-Raposo, and M.~Bessa.
\newblock Presence and cybersickness in immersive content: Effects of content
  type, exposure time and gender.
\newblock {\em Computers \& Graphics}, 71:159--165, 2018.
  \href{https://doi.org/10.1016/j.cag.2017.11.007}
{doi: {{%
10\hspace{.1pt}\discretionary{.}{%
}{.}\hspace{.4pt}1016\discretionary{/}{%
}{/}j\hspace{.1pt}\discretionary{.}{%
}{.}\hspace{.4pt}cag\hspace{.1pt}\discretionary{.}{%
}{.}\hspace{.4pt}2017\hspace{.1pt}\discretionary{.}{%
}{.}\hspace{.4pt}11\hspace{.1pt}\discretionary{.}{%
}{.}\hspace{.4pt}007}}}


\bibitem{Mittelstaedt2019}
J.~M. Mittelstaedt, J.~Wacker, and D.~Stelling.
\newblock Vr aftereffect and the relation of cybersickness and cognitive
  performance.
\newblock {\em Virtual Reality}, 23(2):143–154, jun 2019.
  \href{https://doi.org/10.1007/s10055-018-0370-3}
{doi: {{%
10\hspace{.1pt}\discretionary{.}{%
}{.}\hspace{.4pt}1007\discretionary{/}{%
}{/}s10055\discretionary{%
}{-}{-}018\discretionary{%
}{-}{-}0370\discretionary{%
}{-}{-}3}}}


\bibitem{Nalivaiko2015}
E.~Nalivaiko, S.~L. Davis, K.~L. Blackmore, A.~Vakulin, and K.~V. Nesbitt.
\newblock Cybersickness provoked by head-mounted display affects cutaneous
  vascular tone, heart rate and reaction time.
\newblock {\em Physiology \& Behavior}, 151:583--590, 2015.
  \href{https://doi.org/10.1016/j.physbeh.2015.08.043}
{doi: {{%
10\hspace{.1pt}\discretionary{.}{%
}{.}\hspace{.4pt}1016\discretionary{/}{%
}{/}j\hspace{.1pt}\discretionary{.}{%
}{.}\hspace{.4pt}physbeh\hspace{.1pt}\discretionary{.}{%
}{.}\hspace{.4pt}2015\hspace{.1pt}\discretionary{.}{%
}{.}\hspace{.4pt}08\hspace{.1pt}\discretionary{.}{%
}{.}\hspace{.4pt}043}}}


\bibitem{Nesbitt2017}
K.~Nesbitt, S.~Davis, K.~Blackmore, and E.~Nalivaiko.
\newblock Correlating reaction time and nausea measures with traditional
  measures of cybersickness.
\newblock {\em Displays}, 48:1--8, 2017.
  \href{https://doi.org/10.1016/j.displa.2017.01.002}
{doi: {{%
10\hspace{.1pt}\discretionary{.}{%
}{.}\hspace{.4pt}1016\discretionary{/}{%
}{/}j\hspace{.1pt}\discretionary{.}{%
}{.}\hspace{.4pt}displa\hspace{.1pt}\discretionary{.}{%
}{.}\hspace{.4pt}2017\hspace{.1pt}\discretionary{.}{%
}{.}\hspace{.4pt}01\hspace{.1pt}\discretionary{.}{%
}{.}\hspace{.4pt}002}}}


\bibitem{Nie2020}
G.-Y. Nie, H.~B.-L. Duh, Y.~Liu, and Y.~Wang.
\newblock Analysis on mitigation of visually induced motion sickness by
  applying dynamical blurring on a user's retina.
\newblock {\em IEEE Transactions on Visualization and Computer Graphics},
  26(8):2535--2545, 2020. \href{https://doi.org/10.1109/TVCG.2019.2893668}
{doi: {{%
10\hspace{.1pt}\discretionary{.}{%
}{.}\hspace{.4pt}1109\discretionary{/}{%
}{/}TVCG\hspace{.1pt}\discretionary{.}{%
}{.}\hspace{.4pt}2019\hspace{.1pt}\discretionary{.}{%
}{.}\hspace{.4pt}2893668}}}


\bibitem{Partala}
T.~Partala and V.~Surakka.
\newblock Pupil size variation as an indication of affective processing.
\newblock {\em International Journal of Human-Computer Studies},
  59(1):185--198, 2003.
\newblock Applications of Affective Computing in Human-Computer Interaction.
  \href{https://doi.org/10.1016/S1071-5819(03)00017-X}
{doi: {{%
10\hspace{.1pt}\discretionary{.}{%
}{.}\hspace{.4pt}1016\discretionary{/}{%
}{/}S1071\discretionary{%
}{-}{-}5819\discretionary{%
}{(}{(}03\discretionary{)}{%
}{)}00017\discretionary{%
}{-}{-}X}}}


\bibitem{Peck2020}
K.~Peck, F.~Russo, J.~L. Campos, and B.~Keshavarz.
\newblock Examining potential effects of arousal, valence, and likability of
  music on visually induced motion sickness.
\newblock {\em Experimental Brain Research}, 238(10):2347—2358, October 2020.
  \href{https://doi.org/10.1007/s00221-020-05871-2}
{doi: {{%
10\hspace{.1pt}\discretionary{.}{%
}{.}\hspace{.4pt}1007\discretionary{/}{%
}{/}s00221\discretionary{%
}{-}{-}020\discretionary{%
}{-}{-}05871\discretionary{%
}{-}{-}2}}}


\bibitem{Peterson2020}
R.~A. Peterson and J.~E. Cavanaugh.
\newblock Ordered quantile normalization: a semiparametric transformation built
  for the cross-validation era.
\newblock {\em Journal of Applied Statistics}, 47(13-15):2312--2327, 2020.
  \href{https://doi.org/10.1080/02664763.2019.1630372}
{doi: {{%
10\hspace{.1pt}\discretionary{.}{%
}{.}\hspace{.4pt}1080\discretionary{/}{%
}{/}02664763\hspace{.1pt}\discretionary{.}{%
}{.}\hspace{.4pt}2019\hspace{.1pt}\discretionary{.}{%
}{.}\hspace{.4pt}1630372}}}


\bibitem{Petri2020}
K.~Petri, K.~Feuerstein, S.~Folster, F.~Bariszlovich, and K.~Witte.
\newblock Effects of age, gender, familiarity with the content, and exposure
  time on cybersickness in immersive head-mounted display based virtual
  reality.
\newblock {\em American Journal of Biomedical Sciences}, 12(2), 2020.

\bibitem{R}
{R Core Team}.
\newblock {\em R: A Language and Environment for Statistical Computing}.
\newblock R Foundation for Statistical Computing, Vienna, Austria, 2022.

\bibitem{Radianti2020}
J.~Radianti, T.~A. Majchrzak, J.~Fromm, and I.~Wohlgenannt.
\newblock A systematic review of immersive virtual reality applications for
  higher education: Design elements, lessons learned, and research agenda.
\newblock {\em Computers \& Education}, 147:103778, 2020.
  \href{https://doi.org/10.1016/j.compedu.2019.103778}
{doi: {{%
10\hspace{.1pt}\discretionary{.}{%
}{.}\hspace{.4pt}1016\discretionary{/}{%
}{/}j\hspace{.1pt}\discretionary{.}{%
}{.}\hspace{.4pt}compedu\hspace{.1pt}\discretionary{.}{%
}{.}\hspace{.4pt}2019\hspace{.1pt}\discretionary{.}{%
}{.}\hspace{.4pt}103778}}}


\bibitem{Rebenitsch2021}
L.~Rebenitsch and C.~Owen.
\newblock Estimating cybersickness from virtual reality applications.
\newblock {\em Virtual Reality}, 25(1):165–174, mar 2021.
  \href{https://doi.org/10.1007/s10055-020-00446-6}
{doi: {{%
10\hspace{.1pt}\discretionary{.}{%
}{.}\hspace{.4pt}1007\discretionary{/}{%
}{/}s10055\discretionary{%
}{-}{-}020\discretionary{%
}{-}{-}00446\discretionary{%
}{-}{-}6}}}


\bibitem{Risi2019}
D.~Risi and S.~Palmisano.
\newblock Effects of postural stability, active control, exposure duration and
  repeated exposures on hmd induced cybersickness.
\newblock {\em Displays}, 60:9--17, 2019.
  \href{https://doi.org/10.1016/j.displa.2019.08.003}
{doi: {{%
10\hspace{.1pt}\discretionary{.}{%
}{.}\hspace{.4pt}1016\discretionary{/}{%
}{/}j\hspace{.1pt}\discretionary{.}{%
}{.}\hspace{.4pt}displa\hspace{.1pt}\discretionary{.}{%
}{.}\hspace{.4pt}2019\hspace{.1pt}\discretionary{.}{%
}{.}\hspace{.4pt}08\hspace{.1pt}\discretionary{.}{%
}{.}\hspace{.4pt}003}}}


\bibitem{Sang2006}
F.~D. Y.~P. Sang, J.~P. Billar, J.~F. Golding, and M.~A. Gresty.
\newblock {Behavioral Methods of Alleviating Motion Sickness: Effectiveness of
  Controlled Breathing and a Music Audiotape}.
\newblock {\em Journal of Travel Medicine}, 10(2):108--111, 03 2006.
  \href{https://doi.org/10.2310/7060.2003.31768}
{doi: {{%
10\hspace{.1pt}\discretionary{.}{%
}{.}\hspace{.4pt}2310\discretionary{/}{%
}{/}7060\hspace{.1pt}\discretionary{.}{%
}{.}\hspace{.4pt}2003\hspace{.1pt}\discretionary{.}{%
}{.}\hspace{.4pt}31768}}}


\bibitem{Saredakis2020}
D.~Saredakis, A.~Szpak, B.~Birckhead, H.~A.~D. Keage, A.~Rizzo, and
  T.~Loetscher.
\newblock Factors associated with virtual reality sickness in head-mounted
  displays: A systematic review and meta-analysis.
\newblock {\em Frontiers in Human Neuroscience}, 14, 2020.
  \href{https://doi.org/10.3389/fnhum.2020.00096}
{doi: {{%
10\hspace{.1pt}\discretionary{.}{%
}{.}\hspace{.4pt}3389\discretionary{/}{%
}{/}fnhum\hspace{.1pt}\discretionary{.}{%
}{.}\hspace{.4pt}2020\hspace{.1pt}\discretionary{.}{%
}{.}\hspace{.4pt}00096}}}


\bibitem{Schwind2017}
V.~Schwind, P.~Knierim, C.~Tasci, P.~Franczak, N.~Haas, and N.~Henze.
\newblock {\em "These Are Not My Hands!": Effect of Gender on the Perception of
  Avatar Hands in Virtual Reality}, p. 1577–1582.
\newblock Association for Computing Machinery, New York, NY, USA, 2017.
  \href{https://doi.org/10.1145/3025453.3025602}
{doi: {{%
10\hspace{.1pt}\discretionary{.}{%
}{.}\hspace{.4pt}1145\discretionary{/}{%
}{/}3025453\hspace{.1pt}\discretionary{.}{%
}{.}\hspace{.4pt}3025602}}}


\bibitem{afex}
H.~Singmann, B.~Bolker, J.~Westfall, F.~Aust, and M.~S. Ben-Shachar.
\newblock {\em afex: Analysis of Factorial Experiments}, 2021.
\newblock R package version 1.0-1.

\bibitem{Smith}
S.~P. Smith and S.~Du'Mont.
\newblock Measuring the effect of gaming experience on virtual environment
  navigation tasks.
\newblock In {\em 2009 IEEE Symposium on 3D User Interfaces}, pp. 3--10, 2009.
  \href{https://doi.org/10.1109/3DUI.2009.4811198}
{doi: {{%
10\hspace{.1pt}\discretionary{.}{%
}{.}\hspace{.4pt}1109\discretionary{/}{%
}{/}3DUI\hspace{.1pt}\discretionary{.}{%
}{.}\hspace{.4pt}2009\hspace{.1pt}\discretionary{.}{%
}{.}\hspace{.4pt}4811198}}}


\bibitem{Somrak}
A.~Somrak, M.~Pogačnik, and J.~Guna.
\newblock Suitability and comparison of questionnaires assessing virtual
  reality-induced symptoms and effects and user experience in virtual
  environments.
\newblock {\em Sensors}, 21(4), 2021. \href{https://doi.org/10.3390/s21041185}
{doi: {{%
10\hspace{.1pt}\discretionary{.}{%
}{.}\hspace{.4pt}3390\discretionary{/}{%
}{/}s21041185}}}


\bibitem{STANGLMEIER}
M.~J. Stanglmeier, F.~K. Paternoster, S.~Paternoster, R.~J. Bichler, P.-O.
  Wagner, and A.~Schwirtz.
\newblock Automated driving: A biomechanical approach for sleeping positions.
\newblock {\em Applied Ergonomics}, 86:103103, 2020.
  \href{https://doi.org/10.1016/j.apergo.2020.103103}
{doi: {{%
10\hspace{.1pt}\discretionary{.}{%
}{.}\hspace{.4pt}1016\discretionary{/}{%
}{/}j\hspace{.1pt}\discretionary{.}{%
}{.}\hspace{.4pt}apergo\hspace{.1pt}\discretionary{.}{%
}{.}\hspace{.4pt}2020\hspace{.1pt}\discretionary{.}{%
}{.}\hspace{.4pt}103103}}}


\bibitem{Stanney2020}
K.~Stanney, C.~Fidopiastis, and L.~Foster.
\newblock Virtual reality is sexist: But it does not have to be.
\newblock {\em Frontiers in Robotics and AI}, 7, 2020.
  \href{https://doi.org/10.3389/frobt.2020.00004}
{doi: {{%
10\hspace{.1pt}\discretionary{.}{%
}{.}\hspace{.4pt}3389\discretionary{/}{%
}{/}frobt\hspace{.1pt}\discretionary{.}{%
}{.}\hspace{.4pt}2020\hspace{.1pt}\discretionary{.}{%
}{.}\hspace{.4pt}00004}}}


\bibitem{Stanney2003}
K.~M. Stanney, K.~S. Hale, I.~Nahmens, and R.~S. Kennedy.
\newblock What to expect from immersive virtual environment exposure:
  Influences of gender, body mass index, and past experience.
\newblock {\em Human Factors}, 45(3):504--520, 2003.
\newblock PMID: 14702999. \href{https://doi.org/10.1518/hfes.45.3.504.27254}
{doi: {{%
10\hspace{.1pt}\discretionary{.}{%
}{.}\hspace{.4pt}1518\discretionary{/}{%
}{/}hfes\hspace{.1pt}\discretionary{.}{%
}{.}\hspace{.4pt}45\hspace{.1pt}\discretionary{.}{%
}{.}\hspace{.4pt}3\hspace{.1pt}\discretionary{.}{%
}{.}\hspace{.4pt}504\hspace{.1pt}\discretionary{.}{%
}{.}\hspace{.4pt}27254}}}


\bibitem{Stanney1997}
K.~M. Stanney, R.~S. Kennedy, and J.~M. Drexler.
\newblock Cybersickness is not simulator sickness.
\newblock {\em Proceedings of the Human Factors and Ergonomics Society Annual
  Meeting}, 41(2):1138--1142, 1997.
  \href{https://doi.org/10.1177/107118139704100292}
{doi: {{%
10\hspace{.1pt}\discretionary{.}{%
}{.}\hspace{.4pt}1177\discretionary{/}{%
}{/}107118139704100292}}}


\bibitem{Szpak2019}
A.~Szpak, S.~C. Michalski, D.~Saredakis, C.~S. Chen, and T.~Loetscher.
\newblock Beyond feeling sick: The visual and cognitive aftereffects of virtual
  reality.
\newblock {\em IEEE Access}, 7:130883--130892, 2019.
  \href{https://doi.org/10.1109/ACCESS.2019.2940073}
{doi: {{%
10\hspace{.1pt}\discretionary{.}{%
}{.}\hspace{.4pt}1109\discretionary{/}{%
}{/}ACCESS\hspace{.1pt}\discretionary{.}{%
}{.}\hspace{.4pt}2019\hspace{.1pt}\discretionary{.}{%
}{.}\hspace{.4pt}2940073}}}


\bibitem{Tian2022}
N.~Tian, P.~Lopes, and R.~Boulic.
\newblock A review of cybersickness in head-mounted displays: raising attention
  to individual susceptibility.
\newblock {\em Virtual Reality}, 26(4):1409--1441, Dec 2022.
  \href{https://doi.org/10.1007/s10055-022-00638-2}
{doi: {{%
10\hspace{.1pt}\discretionary{.}{%
}{.}\hspace{.4pt}1007\discretionary{/}{%
}{/}s10055\discretionary{%
}{-}{-}022\discretionary{%
}{-}{-}00638\discretionary{%
}{-}{-}2}}}


\bibitem{Tso}
I.~T.~H. Tso, J.~C.~L. Law, and T.~W.~L. Wong.
\newblock Music-assisted training for dart throwing novices: Post-training
  effects on heart rate and performance accuracy.
\newblock {\em Perceptual and Motor Skills}, 129(1):120--133, 2022.
  \href{https://doi.org/10.1177/00315125211050629}
{doi: {{%
10\hspace{.1pt}\discretionary{.}{%
}{.}\hspace{.4pt}1177\discretionary{/}{%
}{/}00315125211050629}}}


\bibitem{Varmaghani}
S.~Varmaghani, Z.~Abbasi, S.~Weech, and J.~Rasti.
\newblock Spatial and attentional aftereffects of virtual reality and relations
  to cybersickness.
\newblock {\em Virtual Reality}, 26(2):659–668, jun 2022.
  \href{https://doi.org/10.1007/s10055-021-00535-0}
{doi: {{%
10\hspace{.1pt}\discretionary{.}{%
}{.}\hspace{.4pt}1007\discretionary{/}{%
}{/}s10055\discretionary{%
}{-}{-}021\discretionary{%
}{-}{-}00535\discretionary{%
}{-}{-}0}}}


\bibitem{wechsler}
D.~Wechsler.
\newblock Wechsler bellevue adult intelligence scale, 1939.

\bibitem{Weech2020}
S.~Weech, S.~Kenny, M.~Lenizky, and M.~Barnett-Cowan.
\newblock Narrative and gaming experience interact to affect presence and
  cybersickness in virtual reality.
\newblock {\em International Journal of Human-Computer Studies}, 138:102398,
  2020. \href{https://doi.org/10.1016/j.ijhcs.2020.102398}
{doi: {{%
10\hspace{.1pt}\discretionary{.}{%
}{.}\hspace{.4pt}1016\discretionary{/}{%
}{/}j\hspace{.1pt}\discretionary{.}{%
}{.}\hspace{.4pt}ijhcs\hspace{.1pt}\discretionary{.}{%
}{.}\hspace{.4pt}2020\hspace{.1pt}\discretionary{.}{%
}{.}\hspace{.4pt}102398}}}


\bibitem{ggplot}
H.~Wickham.
\newblock {\em ggplot2: Elegant Graphics for Data Analysis}.
\newblock Springer-Verlag New York, 2016.

\bibitem{Wood1988}
C.~D. Wood, J.~E. Manno, M.~J. Wood, B.~R. Manno, and M.~E. Mims.
\newblock Comparison of efficacy of ginger with various antimotion sickness
  drugs.
\newblock {\em Clinical Research Practices and Drug Regulatory Affairs},
  6(2):129--136, 1988.
\newblock PMID: 11538042. \href{https://doi.org/10.3109/10601338809031990}
{doi: {{%
10\hspace{.1pt}\discretionary{.}{%
}{.}\hspace{.4pt}3109\discretionary{/}{%
}{/}10601338809031990}}}


\bibitem{Xie2021}
B.~Xie, H.~Liu, R.~Alghofaili, Y.~Zhang, Y.~Jiang, F.~D. Lobo, C.~Li, W.~Li,
  H.~Huang, M.~Akdere, C.~Mousas, and L.-F. Yu.
\newblock A review on virtual reality skill training applications.
\newblock {\em Frontiers in Virtual Reality}, 2, 2021.
  \href{https://doi.org/10.3389/frvir.2021.645153}
{doi: {{%
10\hspace{.1pt}\discretionary{.}{%
}{.}\hspace{.4pt}3389\discretionary{/}{%
}{/}frvir\hspace{.1pt}\discretionary{.}{%
}{.}\hspace{.4pt}2021\hspace{.1pt}\discretionary{.}{%
}{.}\hspace{.4pt}645153}}}


\end{thebibliography}

\end{document}